\documentclass[12pt]{article}
\usepackage{epsfig}
\def\beq{\begin{equation}}
\def\eeq{\end{equation}}
\def\bc{\begin{center}}
\def\ec{\end{center}}
\def\bea{\begin{eqnarray}}
\def\eea{\end{eqnarray}}

\def\nn{\nonumber}
\def\PR{{\it Phys.~Rev.~}}
\def\PRL{{\it Phys.~Rev.~Lett.~}}
\def\NP{{\it Nucl.~Phys.~}}

\def\PL{{\it Phys.~Lett.~}}

\def\GeV{{\rm GeV}}
\def\gappeq{\mathrel{\rlap {\raise.5ex\hbox{$>$}} {\lower.5ex\hbox{$\sim$}}}}
\def\lappeq{\mathrel{\rlap{\raise.5ex\hbox{$<$}} {\lower.5ex\hbox{$\sim$}}}}

\catcode`@=11
\def\marginnote#1{}
\newcount\hour
\newcount\minute
\newtoks\amorpm
\hour=\time\divide\hour by60
\minute=\time{\multiply\hour by60 \global\advance\minute by-\hour}
\edef\standardtime{{\ifnum\hour<12 \global\amorpm={am}%
        \else\global\amorpm={pm}\advance\hour by-12 \fi
        \ifnum\hour=0 \hour=12 \fi
        \number\hour:\ifnum\minute<10 0\fi\number\minute\the\amorpm}}
\edef\militarytime{\number\hour:\ifnum\minute<10 0\fi\number\minute}
\def\draftlabel#1{{\@bsphack\if@filesw {\let\thepage\relax
   \xdef\@gtempa{\write\@auxout{\string
      \newlabel{#1}{{\@currentlabel}{\thepage}}}}}\@gtempa
   \if@nobreak \ifvmode\nobreak\fi\fi\fi\@esphack}
        \gdef\@eqnlabel{#1}}
\def\@eqnlabel{}
\def\@vacuum{}
\def\draftmarginnote#1{\marginpar{\raggedright\scriptsize\tt#1}}
\def\draft{\oddsidemargin 0.0truein
        \def\@oddfoot{\sl preliminary draft \hfil
        \rm\thepage\hfil\sl\today\quad\militarytime}
        \let\@evenfoot\@oddfoot \overfullrule 3pt
        \let\label=\draftlabel
        \let\marginnote=\draftmarginnote
   \def\@eqnnum{(\theequation)\rlap{\kern\marginparsep\tt\@eqnlabel}%
\global\let\@eqnlabel\@vacuum}  }
\catcode`@=12
\begin{document}
\begin{titlepage}
\vspace*{-1cm}
\phantom{hep-ph/0206077} 
\hfill{DFPD-02/TH/13}
\\
\phantom{hep-ph/0206077} 
\hfill{CERN-TH/2002-127}
\vskip 2.0cm
\begin{center}
{\Large\bf Theoretical Models of Neutrino Masses and Mixings
\footnote{to appear in ``Neutrino Mass'', Springer Tracts in Modern Physics, ed. by G. Altarelli and
K. Winter.}}
\end{center}
\vskip 1.5  cm
\begin{center}
{\large Guido Altarelli}~\footnote{e-mail address: guido.altarelli@cern.ch}
\\
\vskip .1cm
Theory Division, CERN,
\\ 
CH-1211 Gen\`eve 23, Switzerland
\\
\vskip .2cm
{\large Ferruccio Feruglio}~\footnote{e-mail address: feruglio@pd.infn.it}
\\
\vskip .1cm
Dipartimento di Fisica `G.~Galilei', Universit\`a di Padova and
\\ 
INFN, Sezione di Padova, Via Marzolo~8, I-35131 Padua, Italy
\end{center}
\vskip 1.5cm
\begin{abstract}
\noindent
We review theoretical ideas, problems and implications of
different models for neutrino masses and mixing angles. We give a general discussion of
schemes with three or more light neutrinos. Several specific examples
are analyzed in some detail, particularly those that can be embedded into
grand unified theories.
\end{abstract}
\end{titlepage}
\setcounter{footnote}{0}
\vskip2truecm
\section{Introduction}

There is by now convincing evidence, from the experimental study of
atmospheric and solar neutrinos \cite{atmexp,sunexp}, for the
existence of at least two distinct frequencies of neutrino oscillations.
This in turn implies non vanishing neutrino
masses and a mixing matrix, in analogy with the quark sector and the CKM
matrix. So apriori the study of masses and
mixings in the lepton sector should be considered at least as important as that in
the quark sector. But actually there are a number of
features that make neutrinos especially interesting. In fact the smallness
of neutrino masses is probably
related to the fact that
$\nu's$ are completely neutral (i.e. they carry no charge which is exactly
conserved) and are Majorana particles with
masses inversely proportional to the large scale where lepton number (L)
conservation is violated. Majorana masses
can arise from the see-saw mechanism \cite{seesaw}, in which case there is some relation
with the Dirac masses, or from higher-dimensional
non-renormalisable operators which come from a different sector of the
lagrangian density than any other fermion mass terms.
The relation with L non conservation and the fact that the observed
neutrino oscillation frequencies are well compatible
with a large scale for L non-conservation, points to a tantalizing
connection with Grand Unified Theories (Cut's). So
neutrino masses and mixings can represent a probe into the physics at GUT
energy scales and offer a different perspective
on the problem of flavour and the origin of fermion masses. There are also
direct connections with important issues in
astrophysics and cosmology as for example baryogenesis through
leptogenesis \cite{leptgen} and the possibly non-negligible contribution
of neutrinos to hot dark matter in the Universe.

At present there are many alternative models of neutrino masses. This
variety is mostly due to the considerable
experimental ambiguities that still exist. The most crucial questions to
be clarified by experiment are whether the LSND
signal \cite{LSND} will be confirmed or will be excluded and which solar neutrino
solution will eventually be established. If LSND is
right we probably need at least four light neutrinos, 
if not we can do with only the three
known ones. Which solar solution is correct fixes
the corresponding mass squared difference and the associated mixing angle.
Another crucial unknown is the absolute scale of
neutrino masses. This is in turn related to as diverse physical questions
as the possible cosmological relevance of
neutrinos as hot dark matter or the rate of neutrinoless double beta decay ($0\nu\beta\beta$). If
neutrinos are an important fraction of the
cosmological density, say
$\Omega_{\nu}\sim 0.1$, then the average neutrino mass must be
considerably heavier than the splittings that are indicated
by the observed atmospheric and solar oscillation frequencies. For
example, for three light neutrinos, only models with
almost degenerate neutrinos, with common mass
$|m_{\nu}|\approx 1$ eV, are compatible with a large hot dark matter
component, but in this case the existing bounds on
$0\nu\beta\beta$ decay represent an important constraint.  On the contrary
hierarchical three neutrino models (with
both signs of $\Delta m^2_{23}$) have the largest neutrino mass fixed by
$|m|\approx \sqrt{\Delta m^2_{atm}}\approx 0.05$ eV. 
In view of all these
important questions still pending it is no
wonder that many different theoretical avenues are open and have been
explored in the vast literature on the subject.

Here we will briefly summarize the main categories of neutrino mass
models, discuss their respective advantages and
difficulties and give a number of examples. We illustrate how forthcoming
experiments can discriminate among the various
alternatives. We will devote a special attention to the most constrained
set of models, those with only three widely split
neutrinos, with masses dominated by the see-saw mechanism and inversely
proportional to a large mass close to the grand
unification scale
$M_{GUT}$. In this case one can aim at a comprehensive
discussion in a GUT framework of all fermion masses. This is to some
extent possible in models based on SU(5)$\times$
U(1)$_{\rm F}$ or on SO(10) (we always consider SUSY GUT's).

\section{Neutrino Masses and Lepton Number Violation}

Neutrino oscillations imply neutrino masses which in turn demand either the existence of
right-handed (RH) neutrinos (Dirac masses) or lepton number L violation (Majorana masses) or both.
Given that neutrino masses are certainly extremely small, it is really difficult from the theory
point of view to avoid the conclusion that L conservation must be violated. In fact, in terms of
lepton number violation the smallness of neutrino masses can be explained as inversely
proportional to the very large scale where L is violated, of order $M_{GUT}$ or even $M_{Pl}$.

Once we accept L non-conservation we gain an elegant explanation for the smallness of neutrino masses.
If L is not conserved, even in the absence of
heavy RH neutrinos, Majorana masses can be generated for neutrinos by dimension five operators \cite{weinberg} of the form 

\beq 
O_5=\frac{(H l)^T_i \lambda_{ij} (H l)_j}{\Lambda}+~h.c.~~~,
\label{O5}
\eeq 
with $H$ being the ordinary Higgs doublet, $l_i$ the SU(2) lepton doublets, $\lambda$ a matrix in 
flavour space and $\Lambda$ a large scale of mass, of order $M_{GUT}$ or $M_{Pl}$. 
Neutrino masses generated by $O_5$ are of the order
$m_{\nu}\approx v^2/\Lambda$ for $\lambda_{ij}\approx {\rm O}(1)$, where $v\sim {\rm O}(100~\GeV)$ is the
vacuum expectation value of the ordinary Higgs. 

We consider that the existence of RH neutrinos $\nu^c$ is quite plausible because all GUT groups larger than
SU(5) require them. In particular the fact that $\nu^c$ completes the representation 16 of SO(10):
16=$\bar 5$+10+1, so that all fermions of each family are contained in a single representation of
the unifying group, is too impressive not to be significant. At least as a classification group
SO(10) must be of some relevance. Thus in the following we assume that there are both
$\nu^c$ and L non-conservation. With these assumptions the see-saw mechanism \cite{seesaw} is
possible.  Also to fix notations we recall that in its
simplest form it arises as follows. Consider the SU(3) $\times$ SU(2) $\times$ U(1) invariant Lagrangian
giving rise to Dirac and $\nu^c$ Majorana masses (for the time being we consider the $\nu$ Majorana 
mass terms as comparatively negligible):
\beq 
{\cal L}=-{\nu^c}^T y_\nu (H l)+\frac{1}{2} {\nu^c}^T M \nu^c +~h.c.
\label{lag}
\eeq 
The Dirac mass matrix $m_D\equiv y_\nu v/\sqrt{2}$, originating from electroweak symmetry breaking, 
is, in general, non-hermitian and non-symmetric, while the Majorana mass matrix $M$ is symmetric,
$M=M^T$.
We expect the eigenvalues of $M$ to be of order $M_{GUT}$ or more because $\nu^c$ Majorana
masses are SU(3)$\times$ SU(2)$\times$ U(1) invariant, hence unprotected and naturally of the order of the cutoff of the
low-energy theory.  Since all $\nu^c$ are very heavy we can integrate them away.  For this purpose we write down the
equations of motion for $\nu^c$ in the static limit, $i.e.$ neglecting their kinetic terms:
\beq 
-\frac{\partial {\cal L}}{\partial\nu^c}=y_\nu (H l)- M \nu^c= 0~~~.
\label{eulag}
\eeq 
{}From this, by solving for $\nu^c$, we obtain:
\beq
\nu^c= M^{-1} y_\nu (H l)~~~.
\label{R}
\eeq 
We now replace in the lagrangian, eq. (\ref{lag}), this expression for $\nu^c$ and we get
the operator $O_5$ of eq. (\ref{O5}) with
\beq 
\frac{2 \lambda}{\Lambda}=-y_\nu^T M^{-1} y_\nu ~~~~~,
\eeq
and the resulting neutrino mass matrix reads:
\beq 
m_{\nu}=m_D^T M^{-1}m_D~~~.
\eeq 
This is the well known see-saw mechanism result \cite{seesaw}: the light neutrino masses are quadratic in the Dirac
masses and inversely proportional to the large Majorana mass.  If some $\nu^c$ are massless or light they would not be
integrated away but simply added to the light neutrinos. Notice that the above results hold true for any number
$n$ of heavy neutral fermions 
$R$ coupled to the 3 known neutrinos. In this more general case $M$ is an $n$ by $n$ symmetric matrix and the coupling
between heavy and light fields is described by the rectangular $n$ by 3 matrix $m_D$.  Note that for
$m_{\nu}\approx \sqrt{\Delta m^2_{atm}}\approx 0.05$ eV and 
$m_{\nu}\approx m_D^2/M$ with $m_D\approx v
\approx 200~GeV$ we find $M\approx 10^{15}~GeV$ which indeed is an impressive indication for
$M_{GUT}$.

If additional non-renormalisable contributions to $O_5$, eq. \ref{O5}, are comparatively
non-negligible, they should simply be added. After elimination of the heavy right-handed
fields, at the level of the effective low-energy theory, the two types of terms are equivalent. In
particular they have identical transformation properties under a chiral change of basis in flavour
space. The difference is, however, that in the see-saw mechanism, the Dirac matrix
$m_D$ is presumably related to ordinary fermion masses because they are both generated by the Higgs
mechanism and both must obey GUT-induced constraints. Thus if we assume the see-saw mechanism more
constraints are implied. 

\section{Baryogenesis via Leptogenesis from Heavy $\nu^c$ Decay}

In the Universe we observe an apparent excess of baryons over antibaryons. It is appealing that one can
explain the observed baryon asymmetry by dynamical evolution starting from an initial state of the Universe with zero
baryon number (baryogenesis).  For baryogenesis one needs the three famous Sakharov conditions: B violation, CP violation
and no thermal equilibrium. In the history of the Universe these necessary requirements can have occurred at different
epochs. Note however that the asymmetry generated by one epoch could be erased at following epochs if not protected by
some dynamical reason. In principle these conditions could be verified in the SM at the electroweak phase transition. B is
violated by instantons when kT is of the order of the weak scale (but B-L is conserved), CP is violated by the CKM phase
and sufficiently marked out-of- equilibrium conditions could be realized during the electroweak phase transition. So the
conditions for baryogenesis  at the weak scale in the SM superficially appear to be present. However, a more quantitative
analysis
\cite{rev}, shows that baryogenesis is not possible in the SM because there is not enough CP violation and the phase
transition is not sufficiently strong first order, unless
$m_H<80~{\rm GeV}$, which is by now completely excluded by LEP. In SUSY extensions of the SM, in particular in the MSSM, there
are additional sources of CP violations and the bound on $m_H$ is modified by a sufficient amount by the presence of
scalars with large couplings to the Higgs sector, typically the s-top. What is required is that
$m_h\sim 80-110~{\rm GeV}$, a s-top not heavier than the top quark and, preferentially, a small
$\tan{\beta}$. This possibility has by now become very marginal with the results of the LEP2 running.

If baryogenesis at the weak scale is excluded by the data it can occur at or just below the GUT scale, after inflation.
But only that part with
$|{\rm B}-{\rm L}        |>0$ would survive and not be erased at the weak scale by instanton effects. Thus baryogenesis at $kT\sim
10^{10}-10^{15}~{\rm GeV}$ needs B-L violation at some stage like for $m_\nu$ if neutrinos are Majorana particles. The two
effects could be related if baryogenesis arises from leptogenesis then converted into baryogenesis by instantons
\cite{leptgen}. Recent results on neutrino masses are compatible with this elegant possibility \cite{kaol}. Thus the case of
baryogenesis through leptogenesis has been boosted by the recent results on neutrinos \cite{leptog}.

In leptogenesis the departure from equilibrium is determined by the deviation from the average number density
induced by the decay of the heavy neutrinos. The Yukawa interactions of the heavy Majorana neutrinos $\nu^c$
lead to the decays
$\nu^c\rightarrow lH$ (with $l$ a lepton) and $\nu^c\rightarrow \bar l \bar H$ with CP violation. The violation of L conservation
arises from the $\Delta {\rm L}=2$ terms that produce the Majorana mass terms. The rates of the various
interaction processes involved are temperature dependent with different powers of $T$, so that the equilibrium
densities and the temperatures of decoupling from equilibrium during the Universe expansion are different for different
particles and interactions. The rates
$\Gamma_{\Delta L}(T)$ of
$\Delta {\rm L}=2$ processes depend also on the neutrino masses and mixings, so that the observed values of the baryon
asymmetry are related to neutrino processes. Precisely, $\Gamma_{\Delta L}(T)\sim T^3/\Lambda^2$ where $\Lambda$ is the large scale
that appears in eq. (\ref{O5}) and also in the expression of light neutrino masses $m_{\nu}\sim v^2/\Lambda$. The
out-of-equilibrium condition $\Gamma_{\Delta L}(T)< \Gamma_{exp}$, where $\Gamma_{exp}\sim T^2/M_{Pl}$ is the expansion
rate of the Universe, leads to $T\lappeq \Lambda^2/M_{Pl}$ which then implies the relation (when correct proportionality factors
and sum over flavours are included):
\beq
\sum_i m_{\nu_i}^2\lappeq [0.2~ {\rm eV} (\frac{10^{12}~{\rm GeV}}{T})^{1/2}]^2
\label{limit}
\eeq
What exactly is the temperature $T$ which is relevant for leptogenesis depends on the thermal history of the early Universe
and goes beyond the realm of neutrino physics. But if $T\lappeq \Lambda^2/M_{Pl}$ and $\Lambda\sim M_{GUT}$ then the upper limit is
significant and compatible with present neutrino data.

If the RH neutrinos are thermally produced, then the mass of the RH neutrino that drives L violation is limited by
the reheat temperature after inflation, which in turn is typically required not to exceed $10^8-10^{10}$ GeV. 
This limit can be evaded if the RH neutrinos are instead produced by large inflaton oscillations
during the preheating stage \cite{gprt}.

\section{Four (or More) Neutrino Models}

The LSND signal \cite{LSND} has not been confirmed by KARMEN \cite{karmen}. It will be soon
double-checked by MiniBoone \cite{MBoone}. Perhaps it will fade away. But if an oscillation with
$\Delta m^2
\approx 1~ {\rm eV}^2$ is confirmed then, in presence of three distinct frequencies for LSND, atmospheric
\cite{atmexp} and solar \cite{sunexp} neutrino oscillations, the simplest possibility is to introduce at 
least four light
neutrinos. Since LEP has limited to three the number of ``active'' neutrinos (that is with
weak interactions, or equivalently with non-vanishing weak isospin, the only possible gauge charge
of neutrinos) the additional light neutrino(s) $\nu_s$ must be ``sterile'', i.e. with vanishing weak isospin.
Note that $\nu^c$ that appears in the see-saw mechanism, if it exists, is indeed a sterile neutrino, but a
heavy one. 

A possibility to accommodate atmospheric, solar and LSND evidences for neutrino
oscillations without introducing one or more sterile neutrinos is to
invoke CPT violation \cite{cpt}. The required independent frequencies are 
provided by different neutrino and anti-neutrino masses and the fit to
the present data has a good quality \cite{cptfit}.

A typical pattern of masses that works for 4-$\nu$ models consists of two pairs of
neutrinos \cite{4nu} with mass separation between the two pairs, of order $1~{\rm eV}$,
corresponding to the LSND frequency. The upper doublet is almost degenerate at $m^2$ of order
$1~{\rm eV}^2$ being only split by (the mass squared difference corresponding to) the atmospheric (solar)
$\nu$ frequency, while the lower doublet is split by the solar (atmospheric) $\nu$ frequency. 
An alternative to this 2-2 spectrum is given by a 3-1 pattern with 1 being a nearly pure sterile
neutrino separated by the LSND frequency from the 3. The 3-1 spectrum
leads to a comparable (poor) overall quality of fit as the 2-2 pattern.
These mass configurations can be compatible with an important fraction of hot dark matter in the universe. A
complication is that the data appear to be incompatible with pure 2-$\nu$ oscillations for
$\nu_e-\nu_s$ oscillations for solar neutrinos \cite{sunfit} and for $\nu_{\mu}-\nu_s$ oscillations for
atmospheric neutrinos \cite{atmfit}. There are however (after SNO, marginally) viable alternatives. 
One possibility is
obtained by using the large freedom allowed by the presence of 6 mixing angles in the most general
4-$\nu$ mixing matrix. If at least 4 angles are significantly different from zero, one can go beyond pure
2-$\nu$ oscillations and, for example, for solar neutrino oscillations
$\nu_e$ can transform into a mixture of $\nu_a$ and $\nu_s$, where $\nu_a$ is an active neutrino,
itself a superposition of
$\nu_{\mu}$ and $\nu_{\tau}$ (mainly $\nu_{\tau}$) \cite{4nu}. 
A different alternative is to have many interfering sterile
neutrinos: this is the case in the interesting class of models with large extra dimensions, where a whole
tower of Kaluza-Klein neutrinos is introduced. This picture of sterile neutrinos from extra
dimensions appears exciting and we now discuss it in some detail \cite{nuexd}.

The context is theories with large extra dimensions. Gravity propagates in all $D$ dimensions (bulk),
while SM particles live on a 4d brane. As well known \cite{ADD}, this can make the fundamental
scale of gravity $M_D$ much smaller than the Planck mass $M_{Pl}$. In fact for $D~=~\delta~+~4$ we have a 
geometrical factor $V_\delta$, the volume of the compact dimensions, that suppresses gravity, so that
\beq
(M_D)^\delta~V_\delta~=~(M_{Pl}/M_D)^2~~~,
\label{exd1}
\eeq
and, as a result, $M_D$ can be as small as $\sim 1~{\rm TeV}$. 
For neutrino phenomenology we need a really large extra dimension, with a radius $R$ at least of the order
of the scale set by the observed solar oscillation frequencies, $1/R\lappeq 0.01~{\rm eV}$ or $R\gappeq 0.02~mm$.
If we insist on having $M_D$ around $1~{\rm TeV}$ we can assume, for instance, one compact dimension with 
radius $R$ and $\delta-1$ dimensions with a common radius $R'$, such that the volume $V_\delta=
(2\pi)^\delta R R'^{\delta-1}$ 
fits eq. (\ref{exd1}). 
In string theories of gravity there are always scalar fields associated with gravity together with their SUSY
fermionic partners (dilatini, modulini) \cite{bensmi}. These are particles that propagate in the bulk, have no
gauge interactions and can well play the role of sterile neutrinos. The models based on this
framework \cite{modexd,limits} have some good features that make them very appealing at first sight. They
provide a ``physical'' picture for $\nu_s$. In the simplest case the theory includes a 5d fermion $\Psi(x,y)$, 
which decomposes into two 4d Weyl spinors $\nu_s(x,y)$ and $\nu_s'(x,y)$ and contains a KK tower of 
recurrences of sterile neutrinos:
\beq
\nu_s(x,y)~=~\frac{1}{\sqrt{2\pi R}}\sum_n~\nu_s^{(n)}(x) e^{i\frac{ny}{R}}~~~.
\label{4nu1}
\eeq 
The tower mixes with the ordinary light active neutrinos in the lepton
doublet $l$:
\beq 
S_{mix}=\int d^4 x \frac{h}{\sqrt{M_5}} \nu_s(x,0) H(x) l(x)~~~.
\label{4nu2}
\eeq  
The interaction is restricted to the 4d brane at $y=0$ where SM
fields live. Since the 5d spinor $\nu_s(x,y)$ has mass dimension 2, we need the 
mass parameter $M_5\equiv M_D^\delta (2 \pi R')^{\delta-1}$ to keep the Yukawa coupling 
constant dimensionless. From eqs. (\ref{exd1},\ref{4nu1},\ref{4nu2}), after 
electroweak symmetry breaking, we find:
\beq 
S_{mix}=\int d^4 x \sum_n \frac{h v}{\sqrt{2}}\frac{M_D}{M_{Pl}} \nu_s^{(n)}(x) \nu_a(x)~~~,
\label{4nu3}
\eeq  
where $\langle H\rangle \equiv v/\sqrt{2}$ and $\nu_a$ is the active neutrino embedded in $l$. 
Note that the geometrical factor $M_D/M_{Pl}$, which
automatically suppresses the Yukawa coupling $h$, arises naturally from the fact that the sterile
neutrino tower lives in the bulk.
An additional mass parameter $\mu$, related to a possible bulk mass term for the 5d fermion
$\Psi(x,y)$ is also allowed (in more realistic realizations, more 5d fields 
and L-violating interactions can be present). 

The pattern of oscillations results from the superposition of infinite components
with increasing frequencies $\sim n^2$ and decreasing amplitudes $\sim 1/n^2$.
The leading oscillation frequency, $\sqrt{\Delta m^2}$, and the dominant mixing angle 
are determined by $\mu$ and $m=h v M_D/M_{Pl}$, whereas the number of KK excitations that  
effectively take part in the oscillation is controlled by $1/R$. Indeed, if $1/R\gg \sqrt{\Delta m^2}$,
the KK modes, whose masses are approximately given by $n/R$, decouple, with the possible 
exception of the lightest mode. If on the contrary $1/R\lappeq \sqrt{\Delta m^2}$,
then several KK levels participate to the oscillation and the resulting energy dependence
of the survival/conversion probability can appreciably differ from that of the two level case.
Indeed the contribution of a few KK states makes the solar oscillation spectrum more
compatible with the data. Note in passing that $\nu_s$ mixings must be small due to
existing limits from weak processes, supernovae and nucleosynthesis \cite{limits}, so that the preferred solution
for this KK $\nu$ model is MSW SA. Instead 
the KK states should decouple in the case of atmospheric
neutrino oscillations. These constraints fix the range of admissible
values of R as specified above.

In spite of its good properties there
are problems with this picture, in our opinion. The first property that we do not like of models with
large extra dimensions is that the connection with GUT's is lost. In particular the elegant
explanation of the smallness of neutrino masses in terms of the large scale where the L
conservation is violated in general evaporates. Since $M_D\sim 1~{\rm TeV}$ is small, what forbids on the
brane an operator of the form
$(H l)^T_i \lambda_{ij} (H l)_j/M_D$
which would lead to by far too large $\nu$ masses? One
must impose by hand L conservation on the brane and that it is only broken by some Majorana masses
of sterile $\nu$'s in the bulk, which we find somewhat ad hoc. Another problem is that we would
expect gravity to know nothing about flavour, but here we would need RH partners for
$\nu_e$,
$\nu_{\mu}$ and
$\nu_{\tau}$. Also a single large extra dimension has problems, because it implies \cite{an} a
linear evolution of the gauge couplings with energy from $0.01~{\rm eV}$ to $M_D\sim 1~{\rm TeV}$. 
But for more large extra dimensions the
KK recurrences do not decouple fast enough. Perhaps a
compromise at d=2 is possible.
In conclusion the models with large extra dimension are interesting because they are speculative
and fascinating but the more conventional framework still appears more plausible at closer
inspection. 

\section{Three-Neutrino Models}

We now assume that the LSND signal will not be confirmed, so that there are only two distinct neutrino
oscillation frequencies, the atmospheric and the solar frequencies. These two can be reproduced with the
known three light neutrino species (for other reviews of three neutrino models see \cite{us4,barr}). 

Neutrino oscillations are due to a misalignment between the flavour basis, $\nu'\equiv(\nu_e,\nu_{\mu},\nu_{\tau})$, where
$\nu_e$ is the partner of the mass and flavour eigenstate $e^-$ in a left-handed (LH) weak isospin SU(2) doublet (similarly
for 
$\nu_{\mu}$ and $\nu_{\tau})$) and the mass eigenstates $\nu\equiv(\nu_1, \nu_2,\nu_3)$ \cite{pon}: 
\beq
\nu' =U \nu~~~,
\label{U}
\eeq 
where $U$ is the unitary 3 by 3 mixing matrix. Given the definition of $U$ and the transformation properties of the effective
light neutrino mass matrix $m_{\nu}$:
\bea 
\label{tr}
{\nu'}^T m_{\nu} \nu'&= &\nu^T U^T m_\nu U \nu\\ \nonumber 
U^T m_{\nu} U& = &{\rm Diag}\left(m_1,m_2,m_3\right)\equiv
m_{diag}~~~,
\eea 
we obtain the general form of $m_{\nu}$ (i.e. of the light $\nu$ mass matrix in the basis where the charged lepton
mass is a diagonal matrix):
\beq 
m_{\nu}=U m_{diag} U^T~~~.
\label{gen}
\eeq 
The matrix $U$ can be parameterized in terms of three mixing angles $\theta_{12}$,
$\theta_{23}$ and $\theta_{13}$ ($0\le\theta_{ij}\le \pi/2$) 
and one phase $\varphi$ ($0\le\varphi\le 2\pi$) \cite{cab}, exactly as for the quark mixing
matrix $V_{CKM}$. The following definition of mixing angles can be adopted:
\beq 
U~=~ 
\left(\matrix{1&0&0 \cr 0&c_{23}&s_{23}\cr0&-s_{23}&c_{23}     } 
\right)
\left(\matrix{c_{13}&0&s_{13}e^{i\varphi} \cr 0&1&0\cr -s_{13}e^{-i\varphi}&0&c_{13}     } 
\right)
\left(\matrix{c_{12}&s_{12}&0 \cr -s_{12}&c_{12}&0\cr 0&0&1     } 
\right)
\label{ufi}
\eeq 
where $s_{ij}\equiv \sin\theta_{ij}$, $c_{ij}\equiv \cos\theta_{ij}$. 
In addition we have the relative phases among the Majorana masses
$m_1$, $m_2$ and $m_3$. If we choose $m_3$ real and positive, these
phases are carried by $m_{1,2}\equiv\vert m_{1,2} \vert e^{i\phi_{1,2}}$. 
Thus, in general, 9 parameters are added to the SM
when non vanishing neutrino masses are included: 3 eigenvalues, 3 mixing angles and 3 CP violating phases.

In our notation the two frequencies, $\Delta m^2_{I}/4E$ $(I=sun,atm)$, are
parametrized in terms of the $\nu$ mass eigenvalues by 
\beq
\Delta m^2_{sun}\equiv \vert\Delta m^2_{12}\vert ,~~~~~~~
\Delta m^2_{atm}\equiv \vert\Delta m^2_{23}\vert~~~.
\label{fre}
\eeq  
where $\Delta m^2_{12}=\vert m_2\vert^2-\vert m_1\vert^2$ and $\Delta m^2_{23}= m_3^2-\vert m_2\vert ^2$.
The numbering 1,2,3 corresponds to our definition of the frequencies and in principle may not
coincide with the ordering from the lightest to the heaviest state.
\vspace{0.1cm}
\begin{table}[!b]
\caption{Square mass differences and mixing angles \cite{sunfit,atmfit,foli}.
\label{tab01}}
\vspace{0.4cm}
\begin{center}
\begin{tabular}{|c|c|c|c|}   
\hline     
&{\tt lower limit} & {\tt best value} & {\tt upper limit}\\
&($3\sigma$)& & ($3\sigma$)\\
\hline
& & & \\
$(\Delta m^2_{sun})_{\rm LA}~(10^{-5}~{\rm eV}^2)$ & 2.3& 5& 37\\
& & & \\
\hline
& & & \\
$(\Delta m^2_{sun})_{\rm LOW}~(10^{-8}~{\rm eV}^2)$ & 3.5& 8& 12\\
& & & \\
\hline
& & & \\
$\Delta m^2_{atm}~(10^{-3}~{\rm eV}^2)$ & 1& 3& 6\\
& & & \\
\hline
& & & \\
$(\tan^2\theta_{12})_{\rm LA}$ & 0.24 & 0.4 &0.89\\
& & & \\
\hline
& & & \\
$(\tan^2\theta_{12})_{\rm LOW}$ & 0.43 & 0.6 &0.86\\
& & & \\
\hline
& & & \\
$\tan^2\theta_{23}$ & 0.33 & 0.8 &3.3\\
& & & \\
\hline
& & & \\
$\tan^2\theta_{13}$ & 0 & 0 &0.07\\
& & & \\
\hline
\end{tabular} 
\end{center}
\end{table}
{}From experiment, see table \ref{tab01},  we know that $c_{23}\sim s_{23}\sim 1/\sqrt{2}$, corresponding to nearly maximal 
atmospheric neutrino mixing, and that  $s_{13}$ is small, according to CHOOZ, $s_{13}<0.2$ \cite{chooz}. 
The solar angle 
$\theta_{12}$ is probably large (MSW LA, LOW, VO solutions) or even maximal for LOW and VO, but could, 
alternatively, be very small $s_{12}^2\sim O(10^{-3})$ \cite{sunfit}, if the
now disfavoured MSW SA solution is also kept in our list. If we take maximal $s_{23}$ and keep only linear terms in $u= 
s_{13}e^{i\varphi}$ from experiment we find the following structure of the
$U_{fi}$ ($f=e$,$\mu$,$\tau$, $i=1,2,3$) mixing matrix, apart from sign convention redefinitions: 
\beq  U_{fi}= 
\left(\matrix{ c_{12}&s_{12}&u \cr 
-(s_{12}+c_{12}u^*)/\sqrt{2}&(c_{12}-s_{12}u^*)/\sqrt{2}&1/\sqrt{2}\cr
(s_{12}-c_{12}u^*)/\sqrt{2}&-(c_{12}+s_{12}u^*)/\sqrt{2}&1/\sqrt{2}     } 
\right) ~~~~~.
\label{ufi1}
\eeq 
Given the observed frequencies and  our notation in eq. (\ref{fre}), there are three possible patterns of
mass eigenvalues:
\bea
{\rm{Degenerate}}& : & |m_1|\sim |m_2| \sim |m_3|\gg |m_i-m_j|\nonumber\\
{\rm{Inverted~hierarchy}}& : & |m_1|\sim |m_2| \gg |m_3| \nonumber\\
{\rm{Normal~hierarchy} }& : & |m_3| \gg |m_{2,1}|
\label{abc}
\eea 
In the following we will discuss
the phenomenology for these different cases and the respective advantages
and problems.

\subsection{Degenerate Neutrinos}

For degenerate neutrinos the average $m^2$ is much larger than the splittings. At first sight the degenerate case is the
most appealing: the observation of nearly maximal atmospheric neutrino mixing and the experimental indication 
that also the solar
mixing is large (at present the MSW SA solution of the solar neutrino oscillations appears disfavoured by the
data \cite{sunfit}) suggests that all $\nu$ masses are nearly degenerate. Moreover, the common value of
$|m_{\nu}|$ could be compatible with a large fraction of hot dark matter in the universe if
$|m_{\nu}|\sim 1-2$ eV. In this case, however, the existing limits 
\cite{0nubblim} on the absence of 
$0\nu\beta\beta$ ($\vert m_{ee}\vert< 0.2$ eV or to be more conservative $\vert m_{ee}\vert< 0.3-0.5$ eV) imply \cite{gg}
double maximal mixing (bimixing) for solar and atmospheric neutrinos. In fact the quantity which is bound by experiments
is the 11 entry of the
$\nu$ mass matrix, which in general, from eqs. (\ref{tr}) and (\ref{ufi}), is given by :
\beq 
\vert m_{ee}\vert~=\vert(1-s^2_{13})~(m_1 c^2_{12}~+~m_2 s^2_{12})+m_3 e^{2 i\phi} s^2_{13}\vert~~~,
\label{3nu1gen}
\eeq
which in this particular case ($m_3$ cannot compensate for the smallness of $s^2_{13}$) approximately becomes:
\beq 
\vert m_{ee}\vert~\approx~\vert m_1 c^2_{12}~+~m_2 s^2_{12}\vert~\lappeq~0.3-0.5~{\rm eV}~~~.
\label{3nu1}
\eeq 
To satisfy this constraint one needs $m_1\approx -m_2$ (recall that a 
relative phase $\phi_2-\phi_1$ is allowed between $m_1$ and $m_2$) 
and $c^2_{12}\approx s^2_{12}$ to a good accuracy (in
fact we need $\sin^2 2\theta_{12} > 0.96$ in order that
$|\cos2\theta_{12}|~=~|\cos^2\theta_{12}-\sin^2\theta_{12}| < 0.2$). 
This is exemplified by the following texture
\beq
m_\nu
=m
\left(
\begin{array}{ccc}
0& -1/\sqrt{2}& 1/\sqrt{2}\\
-1/\sqrt{2}& (1+\eta)/2& (1+\eta)/2\\
1/\sqrt{2}& (1+\eta)/2& (1+\eta)/2
\end{array}
\right)~~~,
\label{deg3}
\eeq
where $\eta\ll 1$, corresponding to an exact bimaximal mixing, $s_{13}=0$ and the eigenvalues
are $m_1=m$, $m_2=-m$ and $m_3=(1+\eta) m$. This texture has been
proposed in the context of a spontaneously broken SO(3) flavor symmetry
and it has been studied to analyze the stability of the degenerate
spectrum against radiative corrections \cite{BRS}. 
A more realistic mass matrix can be obtained by adding small perturbations to $m_\nu$
in eq. (\ref{deg3}):
\beq
m_\nu
=m
\left(
\begin{array}{ccc}
\delta& -1/\sqrt{2}& (1-\epsilon)/\sqrt{2}\\
-1/\sqrt{2}& (1+\eta)/2& (1+\eta-\epsilon)/2\\
(1-\epsilon)/\sqrt{2}& (1+\eta-\epsilon)/2& 
(1+\eta-2\epsilon)/2
\end{array}
\right)~~~~~~({\tt D1})~~~,
\label{deg4}
\eeq
where $\epsilon$ parametrizes the leading flavor-dependent 
radiative corrections (mainly induced by the $\tau$ Yukawa coupling)
and $\delta$ controls $m_{ee}$. Consider first the case
$\delta\ll \epsilon$. 
To first approximation $\theta_{12}$ remains
maximal. We get $\Delta m^2_{sun}\approx m^2 \epsilon^2/\eta$ 
and
\beq
\theta_{13}\approx 
\left(\frac{\Delta m^2_{sun}}{\Delta m^2_{atm}}
\right)^{{1}/{2}}~~~,~~~~~~~~~
m_{ee}\ll m~\left(\frac{\Delta m^2_{atm}~\Delta m^2_{sun}}{m^4}\right)^
{1/2}~~~.
\label{deg5}
\eeq
If we instead assume $\delta\gg \epsilon$, we find $\Delta m^2_{sun}\approx 
2 m^2 \delta$, $\theta_{23}\approx\pi/4$, $\sin^2 2\theta_{12}\approx 1-\delta^2/4$.
Also in this case the solar mixing angle remains
close to $\pi/4$. We get:
\beq
\theta_{13}\approx 0~~~,~~~~~~~~~~~~~
m_{ee}\approx \frac{\Delta m^2_{sun}}{2 m}~~~,
\label{deg6}
\eeq
too small for detection if the average neutrino mass $m$ is
around the eV scale.
This example shows that there is no guarantee for
$m_{ee}$ to be close to the range of experimental interest, 
even with degenerate neutrinos where the involved masses 
are much larger than the oscillation frequencies.
However, an almost maximal solar mixing angle such as the one implied by
the previous analysis, is difficult to reconcile with the MSW LA solution.   
Of course the strong constraint $s^2_{12}=c^2_{12}$ can be relaxed if the common mass is 
below the hot dark matter maximum. It is true in any case that a signal of
$0\nu\beta\beta$ near the present limit (like a large relic density of hot dark matter) would be an
indication for nearly degenerate
$\nu$'s.

In general, for naturalness reasons, the splittings cannot be too small with respect to
the common mass, unless there is a protective symmetry \cite{BRS,stab}. This is because the wide mass
differences of fermion masses, in particular charged lepton masses, would tend to create neutrino
mass splittings via renormalization group running effects even starting from degenerate masses at a
large scale. For example, for $m\approx 1$ eV, the VO solution for solar neutrino oscillations would
imply $\Delta m/m\sim10^{-9}-10^{-11}$ which is difficult to obtain. 
Even in the previous example, where, for $\delta\ll\epsilon$, the corrections to $\Delta m^2_{sun}$ 
are quadratic in $\epsilon$ rather than linear, we would need $\epsilon<(10^{-3}/m({\rm eV}))^2$ in order
to have $\Delta m^2_{sun}<10^{-9}~{\rm eV}^2$. 
In this respect the MSW LA or LOW solutions would be
favoured, but,  if we insist that $|m_{\nu}|\sim 1-2~{\rm eV}$, it is not clear that the mixing angle 
preferred by the data is sufficiently maximal.
Summarizing, degenerate models with $|m|\sim 1-2~{\rm eV}$ as required if $\nu$'s are a cosmologically important source of
hot dark matter have some problems related to $0\nu\beta\beta$ limits and to naturalness. In
comparison degenerate models with sub-eV common mass appear simpler to realize.

It is clear that in the degenerate case the most likely origin of $\nu$ masses is from some dimension 5
operators $(H l)^T_i\lambda_{ij}(H l)_j/\Lambda$ not related to the see-saw mechanism 
$m_{\nu}=m^T_DM^{-1}m_D$. In fact we expect the $\nu$ Dirac mass $m_D$ not to be degenerate like 
for all other fermions and a conspiracy  to reinstate a
nearly perfect degeneracy between $m_D$ and $M$, which arise from completely different physics,
looks very unplausible (see, however,
\cite{jeza}). Thus in degenerate models, in general, there is no direct relation with
Dirac masses of quarks and leptons and the possibility of a simultaneous description of all fermion
masses within a grand unified theory is more remote \cite{fri}.

The degeneracy of neutrinos
should be guaranteed by some slightly broken symmetry. Models based on discrete or continuous symmetries have been
proposed. For example in the models of ref. \cite{BHKR} the symmetry is SO(3). In the unbroken limit neutrinos are
degenerate and charged leptons are massless. When the symmetry is broken the charged lepton masses are much larger than
neutrino splittings because the former are first order while the latter are second order in the electroweak symmetry
breaking.

A model which is simple to describe but difficult to derive in a natural way is one \cite{fd} where up quarks, down
quarks and charged leptons have ``democratic'' mass matrices, with all entries equal (in first approximation):
\beq 
m_f = {\hat m}_f 
\left(
\matrix{ 1&1&1\cr 
1&1&1\cr 
1&1&1}
\right)+\delta m_f 
\label{dem}~~~,
\eeq 
where ${\hat m}_f$ ($f=u,d,e$) are three overall mass parameters
and $\delta m_f$ denote small perturbations. If we neglect $\delta m_f$, the eigenvalues of $m_f$ are
given by $(0,0,3~{\hat m}_f)$. The mass matrix $m_f$ is diagonalized by a
unitary matrix $U_f$ which is in part determined by the small term $\delta m_f$.
If $\delta m_u\approx \delta m_d$, the CKM matrix, given by $V_{CKM}=U_u^\dagger U_d$, 
is nearly diagonal, due to a compensation between the large mixings contained in 
$U_u$ and $U_d$. When the small terms $\delta m_f$ 
are diagonal and of the form $\delta m_f={\rm Diag}(-\epsilon_f,\epsilon_f,\delta_f)$ the matrices 
$U_f$ are approximately given by (note the analogy with the quark 
model eigenvalues $\pi^0$, $\eta$ and $\eta'$):
\beq 
U_f^\dagger\approx \left(
\matrix{ 1/\sqrt{2}&-1/\sqrt{2}&0\cr 
1/\sqrt{6}&1/\sqrt{6}&-2/\sqrt{6}\cr 
1/\sqrt{3}&1/\sqrt{3}&1/\sqrt{3}}
\right)~~~.
\label{ufri}
\eeq
At the same time, the lightest quarks and charged leptons acquire a non-vanishing mass.
The leading part of the mass matrix in eq. (\ref{dem}) is invariant under a discrete
$S_{3L}\times S_{3R}$ permutation symmetry. The same requirement leads to the general neutrino mass matrix:
\beq
m_{\nu}
=m
\left[
\left(
\begin{array}{ccc}
1& 0& 0\\
0& 1& 0\\
0& 0& 1
\end{array}
\right)
+ r
\left(
\begin{array}{ccc}
1& 1& 1\\
1& 1& 1\\
1& 1& 1
\end{array}
\right)
\right]
+\delta m_{\nu}~~~~~~~~~~({\tt D2})~~~,
\label{deg2}
\eeq
where $\delta m_\nu$ is a small symmetry breaking term and
the two independent invariants are allowed by the Majorana nature of the light neutrinos. If $r$ vanishes the
neutrinos are almost degenerate. In the presence of $\delta m_\nu$ the permutation symmetry is broken
and the degeneracy is removed. 
If, for example, we choose $\delta m_{\nu}={\rm Diag}(0,\epsilon,\eta)$
with $\epsilon<\eta\ll 1$ and $r\ll \epsilon$, the solar and
the atmospheric oscillation frequencies are determined by
$\epsilon$ and $\eta$, respectively. The mixing angles are almost entirely
due to the charged lepton sector. A diagonal $\delta m_{\rm e}$
will lead to a neutrino mixing matrix $U\approx U_e^\dagger$ characterized by
an almost maximal $\theta_{12}$, $\tan^2\theta_{23}
\approx 1/2$ and 
\beq
\theta_{13}\approx \sqrt{m_e/m_\mu}~~~.
\eeq
By going to the basis where the charged leptons are diagonal,
we can see that $m_{ee}$ is close to $m$ and independent from
the parameters that characterize the oscillation phenomena.

The parameter $r$ receives radiative corrections~\cite{fdrad} that, at leading
order, are logarithmic and proportional to the square of the $\tau$ 
lepton Yukawa coupling. It is important to guarantee that this correction 
does not spoil the relation $r\ll \epsilon$, whose violation would lead 
to a completely different mixing pattern. This raises a `naturalness'
problem for the LOW and VO solutions. 
We conclude by stressing that a non-vanishing $\Delta m^2_{atm}$, maximal $\theta_{12}$,
large $\theta_{23}$ and vanishing $\theta_{13}$ are not determined 
by the symmetric limit,
but only by a specific choice of the parameter $r$ and of 
the perturbations that cannot be easily justified on theoretical grounds.
It would be desirable to provide a more sound basis for the choice of
the small terms in this scenario that is quite favourable to
signals both in the $0\nu2\beta$ decay and in sub-leading
oscillations controlled by $\theta_{13}$. 

Anarchical models \cite{anarchy} can be considered as particular cases of degenerate models with
$m^2\sim \Delta m^2_{atm}$. In this class
of models one assumes that all mass matrices are structureless in the
leptonic sector. At present the
data appear to indicate the MSW LA solution as the most likely one. For
this solution the ratio of the solar and
atmospheric frequencies is not so small: typically ${(\Delta m^2_{sun})}_{LA}/\Delta
m^2_{atm}\sim 0.01-0.2$ and two out of
three mixing angles are large. One important observation is that the
see-saw mechanism tends to enhance the ratio
of eigenvalues: it is quadratic in $m_D$ so that a hierarchy factor $f$ in
$m_D$ becomes $f^2$ in $m_\nu$ and the
presence of the Majorana matrix $M$ results in a further widening of the
distribution. Another squaring takes
place in going from the masses to the oscillation frequencies which are
quadratic. As a result a random generation
of the $m_D$ and $M$ matrix elements leads to a distribution of ${(\Delta
m^2_{sun})}_{LA}/\Delta m^2_{atm}$ that peaks around
0.1. At the same time the distribution of $\sin^2{\theta_{ij}}$ is peaked
around 1 for all three mixing angles.
Clearly the smallness of $\theta_{13}$ is problematic. This can be turned
into the prediction that in anarchical
models $\theta_{13}$ must be near the present bound (after all the value
0.2 for $\sin{\theta_{13}}$ is not that
smaller than the maximal value 0.701). In conclusion there is a non
negligible probability that if the MSW LA solution
is realized and $\theta_{13}$ is near the present bound than the neutrino
masses and mixings, interpreted by the
see-saw mechanism, just arise from structureless underlying Dirac and
Majorana matrices.

\subsection{Inverted Hierarchy}

The inverted hierarchy configuration $|m_1|\sim |m_2| \gg |m_3|$ consists of two levels $m_1$ and
$m_2$ with small splitting $\Delta m_{12}^2~=~\Delta m_{sun}^2$ and a common mass given by
$\vert m_{1,2}^2\vert \sim \vert\Delta m_{atm}^2\vert\sim 3\cdot 10^{-3}~eV^2$ (no large hot dark matter component in this
case). One particularly interesting example of this sort \cite{invhier}, which leads to double maximal
mixing, is obtained with the phase choice
$m_1=-m_2$ so that, approximately:
\beq  m_{diag}~=~{\rm Diag} (\sqrt{2} m,-\sqrt{2} m,0)~~~. 
\label{ih1}
\eeq
The effective light neutrino mass matrix
\beq
m_{\nu}~=~Um_{diag}U^T\label{ih2}~~~,
\eeq
which corresponds to the mixing matrix of double maximal mixing $c_{12}=s_{12}=1/\sqrt{2}$   
and $s_{13}=u=0$ in eq. (\ref{ufi1}).
\beq  U_{fi}= 
\left(\matrix{1/\sqrt{2}&1/\sqrt{2}&0 \cr -1/2& 1/2&1/\sqrt{2}\cr 
1/2&-1/2&1/\sqrt{2}     } 
\right) ~~~,
\label{ufi2}
\eeq
is given by:
\beq  m_{\nu}~=~m 
\left(\matrix{ 0&-1&1 \cr -1&0&0\cr
1&0&0     } 
\right) ~~~.
\label{ih3}
\eeq
The structure of $m_{\nu}$ can be reproduced by imposing a flavour symmetry $L_e-L_{\mu}-L_{\tau}$
starting either from $(H l)^T_i\lambda_{ij}(H l)_j/\Lambda$ or from RH neutrinos
via the see-saw mechanism. 
The $1-2$ degeneracy remains stable under radiative
corrections. The preferred solar solutions are VO or the LOW solution. The
MSW LA could be also compatible if the mixing angle is large enough. 

The leading texture in (\ref{ih3}) can be perturbed by adding small terms:
\beq
m_\nu
=m
\left(
\begin{array}{ccc}
\delta& -1& 1\\
-1& \eta& \eta\\
1& \eta& \eta
\end{array}
\right)~~~~~~~~~~({\tt I})~~~,
\label{invh1}
\eeq
where $\delta$ and $\eta$ are small ($\ll 1$), real parameters defined up to 
coefficients of order 1 that can differ in the various matrix elements.
The perturbations leave $\Delta m^2_{atm}$ and $\theta_{23}$ unchanged,
in first approximation. We obtain 
$\tan^2\theta_{12}\approx 1+\delta+\eta$ and 
$\Delta m^2_{sun}/\Delta m^2_{atm}\approx \eta+\delta$, where
coefficients of order one have been neglected.
Moreover $\theta_{13}\approx\eta$. If $\eta\gg\delta$, we
have
\beq
\theta_{13}\approx 
\frac{\Delta m^2_{sun}}{\Delta m^2_{atm}}~~~,~~~~~~~~~~~~~~m_{ee}\ll
\sqrt{\Delta m^2_{sun}} \left(\frac{\Delta m^2_{sun}}{\Delta m^2_{atm}}
\right)^{\frac{1}{2}}
~~~.
\label{ue3ih2}
\eeq
In the other case, $\eta\ll\delta$ we obtain:
\beq
\theta_{13}\ll \frac{\Delta m^2_{sun}}{\Delta m^2_{atm}}~~~,~~~~~~~~~~~~~~
m_{ee}\approx\frac{1}{2}\sqrt{\Delta m^2_{sun}} 
\left(\frac{\Delta m^2_{sun}}{\Delta m^2_{atm}}\right)^{\frac{1}{2}}~~~.
\label{ue3ih1}
\eeq
There is a well-known difficulty of this scenario to fit the 
MSW LA solution~\cite{invhier,bd}. 
Indeed, barring cancellation between
the perturbations, in order to obtain a $\Delta m^2_{sun}$ close to the best fit MSW LA value, 
$\eta$ and $\delta$ should be smaller
than about 0.1 and this keeps the value of $\sin^2 2\theta_{12}$ very close  to 1, 
somewhat in disagreement with global fits of solar data~\cite{sunfit}. 
Even by allowing for a $\Delta m^2_{sun}$ in the upper range of the MSW LA solution,
or some fine-tuning between $\eta$ and $\delta$, we would need
large values of the perturbations to fit the MSW LA solution. 
On the contrary, the LOW solution can be accommodated, but,
in this case, $\theta_{13}$ and $m_{ee}$ estimated in (\ref{ue3ih2},\ref{ue3ih1})
are too small to be detected by planned experiments.

\subsection{Normal Hierarchy}

We now discuss the class of models which we consider of particular interest 
being the
most constrained framework which allows a comprehensive combined study of all fermion
masses in GUT's. We assume three widely split $\nu$'s and the existence of a RH neutrino
for each generation, as required to complete a 16 dimensional representation of SO(10) for each
generation. We then assume dominance of the see-saw
mechanism
$m_{\nu}=m_D^TM^{-1}m_D$. We know that the third-generation eigenvalue of the Dirac mass matrices
of up and down quarks and of charged leptons is systematically the largest one. It is natural
to imagine that this property could also be true for the Dirac mass of $\nu$'s: $m_D^{diag}\sim
{\rm Diag}(0,0,m_{D3})$. After see-saw we expect $m_{\nu}$ to be
even more hierarchical being quadratic in
$m_D$ (barring fine-tuned compensations between $m_D$ and $M$). The amount of hierarchy,
$m^2_3/m^2_2=\Delta m^2_{atm} /\Delta m^2_{sun}$, depends on which solar neutrino solution is
adopted: the hierarchy is maximal for VO and LOW solutions, is moderate for MSW in
general and could become quite mild for the upper $\Delta m^2_{sun}$ domain of the MSW LA
solution. A possible difficulty is that one is used to expect that large splittings correspond
to small mixings because normally only close-by states are strongly mixed. In a 2 by 2 matrix
context the requirement of large splitting and large mixings leads to a condition of vanishing
determinant and large off-diagonal elements. For example the matrix
\beq
\left(\matrix{ x^2&x\cr x&1    } 
\right) 
\label{md000}
\eeq has eigenvalues 0 and $1+x^2$ and for $x$ of O(1) the mixing is large. Thus in the limit of
neglecting small mass terms of order $m_{1,2}$ the demands of large atmospheric neutrino mixing and
dominance of $m_3$ translate into the condition that the 2 by 2 subdeterminant 23 of the 3 by 3
mixing matrix approximately vanishes. The problem is to show that this vanishing can be arranged in
a natural way without fine tuning. Once near maximal atmospheric neutrino mixing is reproduced
the solar neutrino mixing can be arranged to be either small or large without difficulty by
implementing suitable relations among the small mass terms. 

It is not difficult to imagine mechanisms that naturally lead to the approximate vanishing of the
23 sub-determinant. For example \cite{king,king2} assumes that one $\nu^c$ is particularly
light and coupled to
$\mu$ and $\tau$. In a 2 by 2 simplified context if we have
\beq M\propto 
\left(\matrix{ \epsilon&0\cr 0&1    } 
\right)~~~,~~~~~~~ M^{-1}\approx\left(\matrix{ 1/\epsilon&0\cr 0&0    } 
\right)~~~,~~~~~~~m_D=\left(\matrix{a&b\cr c&d} 
\right)~~~,
\label{md0}
\eeq then for a generic $m_D$ we find
\beq m_{\nu}~=~m_D^TM^{-1}m_D\approx 
\frac{1}{\epsilon}\left(\matrix{ a^2&ab\cr ab&b^2   } 
\right)~~~.
\label{md1}
\eeq
A different possibility that we find attractive is that, in the limit of neglecting terms of order
$m_{1,2}$ and, in the basis where charged leptons are diagonal, the Dirac matrix $m_D$,
defined by $\nu^c m_D \nu$, takes the approximate form, called ``lopsided''
\cite{lops1,lopsu1,lops2}:
\beq m_D\propto 
\left(\matrix{ 0&0&0\cr 0&0&0\cr 0&x&1    } 
\right)~~~~~. 
\label{md00}
\eeq This matrix has the property that for a generic Majorana matrix $M$ one finds:
\beq m_{\nu}=m^T_D M^{-1}m_D\propto 
\left(\matrix{ 0&0&0\cr 0&x^2&x\cr 0&x&1    } 
\right)~~~. 
\label{mn0}
\eeq The only condition on $M^{-1}$ is that the 33 entry is non zero. 
However, when the approximately
vanishing matrix elements are replaced by small terms, one must also assume that no new O(1) terms
are generated in $m_{\nu}$ by a compensation between small terms in $m_D$ and large terms in
$M^{-1}$. It is important for the following discussion to observe that
$m_D$ given by eq. (\ref{md00}) under a change of basis transforms as $m_D\to V^{\dagger} m_D U$
where $V$ and $U$ rotate the right and left fields respectively. It is easy to check that in order to
make $m_D$ diagonal we need large left mixings (i.e. large off diagonal terms in the matrix that
rotates LH fields).
Thus the question is how to reconcile large LH mixings
in the leptonic sector with the observed near diagonal form of $V_{CKM}$, the quark mixing matrix.
Strictly speaking, since $V_{CKM}=U^{\dagger}_u U_d$, the individual matrices $U_u$ and $U_d$ need
not be near diagonal, but
$V_{CKM}$ does, while the analogue for leptons apparently cannot be near diagonal. However for quarks nothing
forbids that, in the basis where $m_u$ is diagonal, the $d$ quark matrix has large non
diagonal terms that can be rotated away by a pure RH rotation. We suggest that this is so
and that in some way RH mixings for quarks correspond to LH mixings for leptons.

In the context of (SUSY) SU(5) there is a very attractive hint of how the present
mechanism can be realized \cite{lopsu5af,lopsu5}. In the
$\bar 5$ of SU(5) the $d^c$ singlet appears together with the lepton doublet $(\nu,e)$. The $(u,d)$
doublet and $e^c$ belong to the 10 and $\nu^c$ to the 1 and similarly for the other families. As a
consequence, in the simplest model with mass terms arising from only Higgs pentaplets, the Dirac
matrix of down quarks is the transpose of the charged lepton matrix:
$m_d=(m_l)^T$. Thus, indeed, a large mixing for RH down quarks corresponds to a large
LH mixing for charged leptons.  At leading order we may have
the lopsided texture:
\beq 
m_d=(m_l)^T=
\left(
\matrix{ 0&0&0\cr 0&0&1\cr 0&0&1}
\right) v_d~~~.
\eeq 
In the same simplest approximation with  5 or $\bar 5$ Higgs, the up quark mass matrix is
symmetric, so that left and right mixing matrices are equal in this case. Then small mixings for up
quarks and small LH mixings for down quarks are sufficient to guarantee small $V_{CKM}$
mixing angles even for large $d$ quark RH mixings.  
It is well known that a model where the down
and the charged lepton matrices are exactly the transpose of one another cannot be exactly true
because of the $e/d$ and
$\mu/s$ mass ratios. It is also known that one remedy to this problem is to add some Higgs
component in the 45 representation of SU(5) \cite{jg}. But the symmetry under transposition can still be a good guideline if we are
only interested in the order of magnitude of the matrix entries and not in their exact values.
Similarly, the Dirac neutrino mass matrix
$m_D$ is the same as the up quark mass matrix in the very crude model where the Higgs pentaplets
come from a pure 10 representation of SO(10):
$m_D=m_u$. For $m_D$ the dominance of the third family eigenvalue  as well as a near diagonal
form could be an order of magnitude remnant of this broken symmetry. Thus, neglecting small terms,
the neutrino Dirac matrix in the basis where charged leptons are diagonal could be directly
obtained in the form of eq. (\ref{md00}).

To get a realistic mass matrix, we allow for
deviations from the symmetric limit in (\ref{mn0}), where we take $x=1$. For instance, we can consider those models where
the neutrino mass matrix elements are dominated,
via the see-saw mechanism, by the exchange of two right-handed
neutrinos~\cite{king2}. Since the exchange of a single RH neutrino
gives a successful zeroth order texture, we are encouraged to
continue along this line.
Thus, we add a sub-dominant contribution of a second
RH neutrino,
assuming that the third one
gives a negligible contribution
to the neutrino mass matrix, because
it has much smaller Yukawa couplings
or is much heavier than the first two.
The Lagrangian
that describes this plausible subset of see-saw models,
written in the mass eigenstate
basis of RH neutrinos and charged leptons,
is
\beq
{\cal L} = y_i \nu^c H l_i + y'_i {\nu^c}' H l_i +
\frac{M}{2} {\nu^c}^2 + \frac{M'}{2} {\nu^c}^{\prime 2}~~~,
\eeq
leading to
\beq
({m_\nu})_{ij} \propto
\frac{y_i y_j}{M} +
\frac{y'_i y'_j}{M'}~~~,
\eeq
where $i,j=\{e,\mu,\tau\}$.
In particular, if $y_e\ll y_\mu\approx y_\tau$ and $y'_\mu\approx
y'_\tau$, we obtain:
\beq
m_\nu
=m
\left(
\begin{array}{ccc}
\delta& \epsilon& \epsilon\\
\epsilon& 1+\eta& 1+\eta\\
\epsilon& 1+\eta& 1+\eta
\end{array}
\right)
\label{hier1}~~~~~~~~~~({\tt N})~~~,
\eeq
where coefficients of order one multiplying the small quantities 
$\delta$, $\epsilon$ and $\eta$ have been omitted.
The mass matrix in (\ref{hier1}) does not describe the most
general perturbation of the zeroth order texture  (\ref{mn0}).
We have implicitly assumed a symmetry between $\nu_\mu$ 
and $\nu_\tau$ which is preserved by the perturbations,
at least at the level of the order of magnitudes.
The perturbed texture (\ref{hier1}) can also arise
when the zeroes of the lopsided Dirac matrix in (\ref{md00})
are replaced by small quantities. 
It is possible to construct models along this line based
on a spontaneously broken U(1)$_{\rm F}$ flavor symmetry, where  
$\delta$, $\epsilon$ and $\eta$ are given by positive powers
of one or more symmetry breaking parameters. Moreover, by playing
with the U(1)$_{\rm F}$ charges, we can adjust, to certain extent, 
the relative hierarchy between $\eta$, $\epsilon$ and 
$\delta$~\cite{king,king2,lopsu1,lops2,lopsu5af,lopsu5}, as we will see in section 8. The texture (\ref{hier1}) can also be generated in SUSY models
with $R$-parity violation \cite{rpv}. 

After a first rotation by an angle $\theta_{23}$ close to 
$\pi/4$ and a second rotation with $\theta_{13}\approx \epsilon$, we get
\beq
m_\nu
\approx m
\left(
\begin{array}{ccc}
\delta+\epsilon^2& \epsilon&0 \\
\epsilon& \eta& 0\\
0& 0& 2
\end{array}
\right)
\label{hier2}~~~,
\eeq
up to order one coefficients in the small entries.
To obtain a large solar mixing angle, we need $|\eta-\delta|<
\epsilon$. In realistic models there is no reason for a cancellation 
between independent perturbations and thus we assume both
$\delta\le\epsilon$ and $\eta\le\epsilon$.  

Consider first the case $\delta\approx \epsilon$ and $\eta<\epsilon$.
The solar mixing angle $\theta_{12}$ is large
but not maximal, as preferred by the MSW LA solution.
We also have $\Delta m^2_{atm}\approx 4 m^2$, $\Delta m^2_{sun}
\approx \Delta m^2_{atm} \epsilon^2$ and 
\beq
m_{ee}\approx \sqrt{\Delta m^2_{sun}}~~~.
\label{bestnh}
\eeq

If $\eta\approx \epsilon$ and $\delta\ll\epsilon$,  
we still have a large solar mixing angle and
$\Delta m^2_{sun}\approx \epsilon^2\Delta m^2_{atm}$, as before.
However $m_{ee}$ will be much smaller than the estimate in (\ref{bestnh}).
Unfortunately, this is the case of the models based on the
above mentioned U(1)$_{\rm F}$ flavor symmetry that, at least in its
simplest realization, tends to predict $\delta\approx \epsilon^2$.
In this class of models we find 
\beq
m_{ee}\approx \sqrt{\Delta m^2_{sun}}
\left(
\frac{\Delta m^2_{sun}}{\Delta m^2_{atm}}
\right)^{\frac{1}{2}}~~~, 
\label{nh2}
\eeq
below the sensitivity of the next generation of planned 
experiments. It is worth to mention that in both cases
discussed above, we have 
\beq
\theta_{13}\approx 
\left(\frac{\Delta m^2_{sun}}{\Delta m^2_{atm}}\right)^{\frac{1}{2}}~~~,
\label{ue3nh}
\eeq
which might be very close to the present experimental 
limit.

If both $\delta$ and $\eta$ are much smaller than
$\epsilon$, the 12 block of $m_\nu$ has an approximate pseudo-Dirac
structure and the angle $\theta_{12}$ becomes maximal.
This situation is typical of some models where leptons 
have U(1)$_{\rm F}$ charges
of both signs whereas the order parameters of U(1)$_{\rm F}$ breaking
have all charges of the same sign~\cite{lopsu5af}. 
We have two eigenvalues approximately given by $\pm m~\epsilon$.
As an example, we consider the case where $\eta=0$ and 
$\delta\approx\epsilon^2$. We find $\sin^2 2\theta_{12}\approx
1-\epsilon^2/4$, $\Delta m^2_{sun}\approx m^2 \epsilon^3$
and
\beq
\theta_{13}\approx 
\left(\frac{\Delta m^2_{sun}}{\Delta m^2_{atm}}\right)^{\frac{1}{3}}~~~,
~~~~~~~~~~~
m_{ee}\approx \sqrt{\Delta m^2_{sun}}
\left(
\frac{\Delta m^2_{sun}}{\Delta m^2_{atm}}
\right)^{\frac{1}{6}}~~~. 
\label{nh3}
\eeq
In order to recover the MSW LA solution we would
need a relatively large value of $\epsilon$. This is in general
not acceptable because, on the one hand the presence of
a large perturbation raises doubts about the consistency 
of the whole approach and, on the other hand, in existing models
where all fermion sectors are related to each other, $\epsilon$
is never larger than the Cabibbo angle. 
Therefore, the case $\delta,\eta\ll\epsilon$ can be more easily adapted to fit the LOW solution where the solar frequency is small. As a consequence,
$m_{ee}$ is beyond the reach of the next generation of experiments,
whereas $\theta_{13}$ might be tested at future facilities.

\subsection{Summary}

Given the present experimental knowledge, which favours $\Delta m^2_{atm}$ as 
the leading oscillation frequency and two large mixing angles, $\theta_{23}$ and $\theta_{12}$, it is natural to define a
zeroth order approximation of the theory, where $\Delta m^2_{sun}$
and $\theta_{13}$ vanish (which allows us to neglect the CP-breaking 
parameter $\varphi$) whereas $\theta_{23}$ and $\theta_{12}$ are maximal.
For each pattern of neutrino masses, we have considered
the most interesting textures that arise in this
limit. 
\vspace{0.1cm}
\begin{table}[!h]
\caption{Order-of-magnitude estimates for $\theta_{13}$.
\label{tab1}}
\vspace{0.4cm}
\begin{center}
\begin{tabular}{|c|c|c|c|}   
\hline                         
& & &  \\
{\tt Texture}& $\theta_{12}$ & $\theta_{13}$ &  
{\tt Perturbations}\\
& & &  \\
\hline
& & &  \\
& 
& 
$\approx \left(\frac{\Delta m^2_{sun}}{\Delta m^2_{atm}}\right)^{1/2}$&
$\epsilon\gg\delta$\\
$m_\nu^{\rm D1}=
m
\left(
\begin{array}{ccc}
\delta& -\frac{1}{\sqrt{2}}& \frac{(1-\epsilon)}{\sqrt{2}}\\
-\frac{1}{\sqrt{2}}& \frac{(1+\eta)}{2}& \frac{(1+\eta-\epsilon)}{2}\\
\frac{(1-\epsilon)}{\sqrt{2}}& \frac{(1+\eta-\epsilon)}{2}& 
\frac{(1+\eta-2\epsilon)}{2}
\end{array}
\right)$
&
$\approx \pi/4$&
&
\\
&
&
$\approx 0$&
$\epsilon\ll\delta$\\
& & &  \\
\hline                        
& & &  \\
$m_\nu^{\rm D2}=
m
\left[
\left(
\begin{array}{ccc}
1& 0& 0\\
0& 1& 0\\
0& 0& 1
\end{array}
\right)
+ r
\left(
\begin{array}{ccc}
1& 1& 1\\
1& 1& 1\\
1& 1& 1
\end{array}
\right)
\right]
+\delta m_{\nu}$
&
$\approx \pi/4$&
$\approx \left(\frac{m_e}{m_\mu}\right)^{1/2}$&
\\
& & & \\
\hline                        
& & & \\
& 
& 
$\approx \frac{\Delta m^2_{sun}}{\Delta m^2_{atm}}$&
$\eta\gg\delta$\\
$m_\nu^{\rm I}=
m
\left(
\begin{array}{ccc}
\delta& -1& 1\\
-1& \eta& \eta\\
1& \eta& \eta
\end{array}
\right)$
&
$\approx \pi/4$&
&
\\
& 
& 
$\ll \frac{\Delta m^2_{sun}}{\Delta m^2_{atm}}$&
$\eta\ll\delta$\\
& & & \\
\hline
& & & \\
& & & 
$\eta<\delta\approx\epsilon$\\
&
$O(1)$&
$\approx \left(\frac{\Delta m^2_{sun}}{\Delta m^2_{atm}}\right)^{1/2}$&
\\
$m_\nu^{\rm N}=
m
\left(
\begin{array}{ccc}
\delta& \epsilon& \epsilon\\
\epsilon& 1+\eta& 1+\eta\\
\epsilon& 1+\eta& 1+\eta
\end{array}
\right)$
& & & 
$\delta\approx\epsilon^2\approx\eta^2$\\
& & & \\
&
$\approx \pi/4$&
$\approx \left(\frac{\Delta m^2_{sun}}{\Delta m^2_{atm}}\right)^{1/3}$&
$\delta\approx\epsilon^2~~~\eta=0$\\
& & & \\
\hline
\end{tabular} 
\end{center}
\end{table}
This approximation is of course not realistic and should be regarded
only as a limiting case, possibly arising from an underlying symmetry.
Many effects can perturb this limit, such as small symmetry breaking
terms, radiative corrections, effects coming from residual
rotations needed to diagonalize the charged lepton mass matrix, or
to render canonical the leptonic kinetic terms. Some of these
will be discussed more in detail in the next sections. 
It turns out that in most of the existing models the leading textures 
are modified by small perturbations having a 
simple structure, such as those called {\tt D1}, {\tt D2}, {\tt I}
and {\tt N} in eqs. (\ref{deg4}), (\ref{deg2}), (\ref{invh1}) and
(\ref{hier1}).

We have analyzed
these perturbations, with the hope that the results
are sufficiently representative of the many existing models.
Of course there is no guarantee that this discussion can
cover all theoretical possibilities. Moreover, if 
$\Delta m^2_{sun}$ and $\theta_{13}$ 
were as large as experimentally allowed, the perturbations 
would become large, the whole approach could become questionable
and the data would be more appropriately described by
an anarchical framework.

A remarkable feature is that most models continue to predict 
an almost maximal
solar angle, even after inclusion of the 
perturbations. This is often due to an approximate pseudo-Dirac
structure in the 12 sector, which, at leading order,
forces $\theta_{12}=\pi/4$. Exceptions to this trend
are given by some of the possibilities offered by the 
normal hierarchy, for which $\theta_{12}$ is undetermined
at leading order. 

It is also apparent that, apart from the case of degenerate
spectrum realized with a texture similar to the one of flavor democracy
(D2), the possibility of measuring $m_{ee}$
with the next generation of experiments seems to be significant
only if the solar oscillation frequency is very close to the
upper part of the range allowed by the MSW LA solution.

\section{Importance of Neutrinoless Double Beta Decay}

The discovery of $0\nu \beta \beta$ decay would be very important because it would establish lepton number violation and
the Majorana nature of $\nu$'s. Indeed oscillation experiments cannot distinguish between pure Dirac and Majorana
neutrinos. Moreover, the search for $0\nu \beta\beta$ decay provides information about the absolute
spectrum, while neutrino oscillations are only sensitive to mass differences.  
Complementary information on the sum of neutrino masses is also provided
by the galaxy power spectrum combined with measurements of the cosmic
microwave background anisotropies \cite{galaxies}.
As already mentioned the present limit from $0\nu \beta \beta$ is $\vert m_{ee}\vert< 0.2$ eV or to be more conservative
$\vert m_{ee}\vert < 0.3-0.5$ eV \cite{0nubblim}. 
In this respect it is interesting to see what is the level at which a signal can be expected or at
least not excluded in the different classes of models \cite{fsv,0nubb}. For 3-neutrino models with degenerate, inverse hierarchy or normal
hierarchy mass patterns, starting from the general formula in eq. (\ref{3nu1gen}), 
it is simple to derive the following bounds.
\begin{itemize}
\item[a)] 
Degenerate case. If $|m|$ is the common mass, apart from a phase, and taking $s_{13}=0$, which, 
as already observed, is a safe approximation in this case, we have $m_{ee}=|m|(c_{12}^2\pm s_{12}^2)$. 
Here the phase ambiguity has been reduced to a sign ambiguity which is sufficient for deriving bounds. 
So, depending on the sign we have $m_{ee}=|m|$ or $m_{ee}=|m|cos2\theta_{12}$. We conclude that in 
this case $m_{ee}$ could be very close to the present experimental limit
because $|m|$ can be sizeably larger than the $0\nu\beta\beta$ bound ($|m|< O(1 {\rm eV})$), but should be at least of 
order $O(\sqrt{\Delta m^2_{atm}})~\sim~O(10^{-2}~ {\rm eV})$ unless the solar angle is practically maximal, 
in which case the minus sign option can be as small as required.
\item[b)] 
Inverse hierarchy case. In this case the same approximate formula $m_{ee}=|m|(c_{12}^2\pm s_{12}^2)$ holds 
because $m_3$ is
small and $s_{13}$ can be neglected. The difference is that here we know that $|m|\approx 
\sqrt{\Delta m^2_{atm}}$ so that $\vert m_{ee}\vert<\sqrt{\Delta
m^2_{atm}}~\sim~0.05$ eV.
\item[c)] 
Normal hierarchy case. Here we cannot in general neglect the $m_3$ term. However in this case $\vert m_{ee}\vert~\sim~
\sqrt{\Delta m^2_{sun}}~ s_{12}^2~+~\sqrt{\Delta m^2_{atm}}~ s_{13}^2$ and we have the bound 
$\vert m_{ee}\vert <$ a few $10^{-3}$ eV.
\end{itemize}

Recently evidence for $0\nu \beta \beta$ was claimed in ref. \cite{kla} at the 2-3$\sigma$ level with
$\vert m_{ee}\vert\sim~0.39~{\rm eV}$. If confirmed this would rule out cases b) and c) and point to case a) or to models with more than
3 neutrinos.

\section{Expectations for $\theta_{13}$}

The measurement of $\theta_{13}$ represents one of the main challenges
for the next generations of experiments on neutrino oscillations,
which, with the help of very intense neutrino beams \cite{inb}, might reach a 
sensitivity of few percent on $\theta_{13}$. A sizeable
$\theta_{13}$ would have an important impact on the observability
of CP-violating effects in the leptonic sector. We collect in table \ref{tab1}
our estimates of $\theta_{13}$ for the various textures considered
in section 5. 
\begin{figure}[!h]
\includegraphics[height=3.0in]{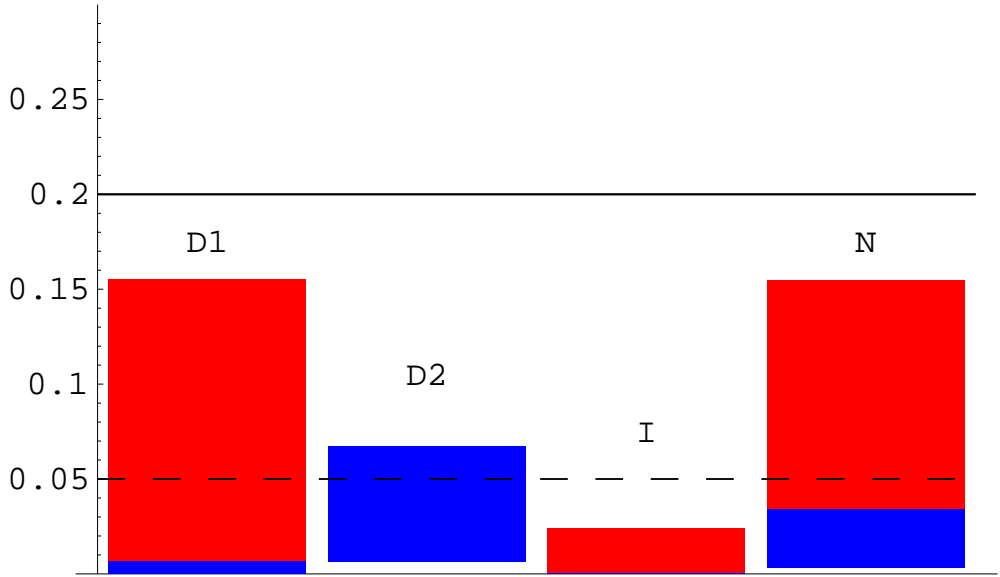}
\caption{Order-of-magnitude estimates of $\theta_{13}$, for MSW LA solution (red) and 
LOW solution (blu). There is no difference between MSW LA and LOW
for the texture D2. The upper limits have been obtained
by using the approximate 
expression of table 1 and the values: 
$\Delta m^2_{atm}=2.5\cdot 10^{-3}~eV^2$,  $\Delta m^2_{sun}=
6.0\cdot 10^{-5}~eV^2~(10^{-7}~eV^2)$ for MSW LA (LOW).   
The continuous and dashed lines show the present upper bound
on $\theta_{13}$ and the possible reach of long baseline
experiments with high intensity neutrino beams.
\label{profiles}}
\end{figure}
Figure \ref{profiles} displays the expectations for $\theta_{13}$. 
Such expectations are of course very
rough and are not meant to be statistically meaningful.
It is however interesting to note that the MSW LA solution favours $\theta_{13}$
in an experimentally accessible range, for all textures but for the 
inverted hierarchy, where we find a strong suppression.
The LOW solution prefers a smaller, unobservable $\theta_{13}$,
with the possible exception of the texture D2, corresponding to 
flavour democracy. We recall that if the neutrino mass matrix
is structureless as advocated by the anarchical framework, then
$\theta_{13}$ is naturally expected to be very close to its present 
experimental bound.

\section{Grand Unified Models of Fermion Masses}

We have seen that the smallness of neutrino masses interpreted via the see-saw mechanism directly leads to a scale $\Lambda$ 
for L non-conservation which is remarkably close to $M_{GUT}$. Thus neutrino masses and mixings should find a
natural context in a GUT treatment of all fermion masses. The hierarchical pattern of quark and lepton masses, within
a generation and across generations, requires some dynamical suppression mechanism that acts differently on the
various particles. 
This hierarchy can be generated by a number of operators of different dimensions suppressed by inverse
powers of the cut-off $\Lambda_c$ of the theory. In some realizations, the different powers of $1/\Lambda_c$
correspond to different orders in some symmetry breaking parameter $v_f$ arising from the spontaneous 
breaking of a flavour symmetry.
In the next subsections we describe some simplest models based
on SU(5) $\times$ U(1)$_{\rm F}$ and on SO(10) which illustrate these possibilities.

\subsection{Models Based on Horizontal Abelian Charges}
 
We discuss here some explicit examples of grand unified models in the framework of a unified SUSY
SU(5) theory with an additional 
U(1)$_{\rm F}$ flavour symmetry. The SU(5) generators  act ``vertically'' inside one generation, while the
U(1)$_{\rm F}$ charges are different ``horizontally'' from one generation to the other. If, for a
given interaction vertex, the
U(1)$_{\rm F}$ charges do not add to zero, the vertex is forbidden in the symmetric limit. But the symmetry is spontaneously broken by the VEV's
$v_f$ of a number of ``flavon'' fields with non vanishing charge. Then a forbidden coupling is rescued but is
suppressed by powers of the small parameters $v_f/\Lambda_c$ with the exponents larger for larger charge 
mismatch \cite{u1}. We expect
$v_f\gappeq M_{GUT}$ and
$\Lambda_c \lappeq M_{Pl}$. Here we discuss some aspects of the description of fermion masses in this framework.

In these models the known generations of quarks and leptons are contained in triplets
$\Psi^{10}_i$ and
$\Psi^{\bar 5}_i$, $(i=1,2,3)$ corresponding to the 3 generations, transforming as $10$ and ${\bar 5}$ of SU(5),
respectively. Three more
SU(5) singlets
$\Psi^1_i$ describe the RH neutrinos. In SUSY models we have two Higgs multiplets, which transform as 5
and $\bar 5$ in the minimal model. The two Higgs multiplets may have the same or different charges. We can arrange
the unit of charge in such a way that the Cabibbo angle, which we consider as the typical hierarchy parameter of
fermion masses and mixings, is obtained when the suppression exponent is unity. Remember that the Cabibbo angle is
not too small, $\lambda \sim 0.22$ and that in U(1)$_{\rm F}$ models all mass matrix elements are of the form of a power
of a suppression factor times a number of order unity, so that only their order of suppression is defined. As a
consequence, in practice,  we can limit ourselves to integral charges in our units (for example,
$\sqrt{\lambda} \sim 1/2$ is already almost unsuppressed). 

There are many variants of these models: fermion charges can all be non negative with only negatively charged
flavons, or there can be fermion charges of different signs with either flavons of both charges or only flavons of
one charge. We can have that only the top quark mass is allowed in the symmetric limit, or that also other third
generation fermion masses are allowed. The Higgs charges can be equal, in particular both vanishing or can be
different. We can arrange that all the structure is in charged fermion masses while neutrinos are anarchical.  

\subsubsection{F(fermions)$\ge$ 0}

Consider, for example, a simple model with all charges of matter fields being non negative and containing one
single flavon ${\bar\theta}$ of charge F$=-1$. For a maximum of simplicity we also assume that all the  third generation masses
are directly allowed in the symmetric limit. This is realized by taking vanishing charges for the Higgses and for
the third generation components $\Psi^{10}_3$, $\Psi^{\bar 5}_3$ and $\Psi^1_3$. For example, 
if we define F$(\Psi^R_i)\equiv q^R_i$ $(R=10,{\bar 5},1;~i=1,2,3)$
we could take
\cite{chargesnu,lopsu1,lops2,lopsu5af} (see also
\cite{u1again})
\bea
(q^{10}_1,q^{10}_2,q^{10}_3) &= & (3,2,0) \nn\\
(q^{\bar 5}_1,q^{\bar 5}_2,q^{\bar 5}_3) &= & (2,0,0)~~~. 
\label{cha}
\eea        
A generic mass matrix has the form
\beq 
m~=~
\left(
\matrix{ y_{11}\lambda^{q_1+q'_1}&y_{12}\lambda^{q_1+q'_2}&y_{13}\lambda^{q_1+q'_3}\cr
y_{21}\lambda^{q_2+q'_1}&y_{22}\lambda^{q_2+q'_2}&y_{23}\lambda^{q_2+q'_3}\cr
y_{31}\lambda^{q_3+q'_1}&y_{32}\lambda^{q_3+q'_2}&y_{33}\lambda^{q_3+q'_3}}
\right) v~~~,
\label{m1}
\eeq  
where all the $y_{ij}$ are of order 1 and $(q_i,q'_j)$ are the charges of $(\Psi^{10},\Psi^{10})$ for $m_u$, of
$(\Psi^{\bar 5},\Psi^{10})$ for $m_d$ or $m^T_l$, of $(\Psi^{1},\Psi^{\bar 5})$ for $m_D$ (the Dirac $\nu$ mass), and of $(\Psi^{1},\Psi^{1})$ for $M$, the Majorana
$\nu^c$ mass, respectively. We have $\lambda\equiv\langle{\bar \theta}\rangle/\Lambda_c$ and the 
quantity $v$ represents the appropriate VEV or mass parameter.
It is important to observe that $m$ can be written as:
\beq 
m~=~\lambda_R y \lambda_{R'} v~~~,
\label{m1p}
\eeq    
where $\lambda_R={\rm Diag}(\lambda^{q_1^R},\lambda^{q_2^R},\lambda^{q_3^R})$ and $y$ is the $y_{ij}$ matrix. The models
with all non negative charges and one single flavon have particularly simple factorization properties. For example,
if we start from the Dirac
$\nu$ matrix: $m_D~=~\lambda_1 y_D \lambda_{\bar 5} v_u$ and the $\nu^c$ Majorana matrix 
$M~=~\lambda_1 y_M \lambda_1 \Lambda$ and write down the see-saw expression for $m_{\nu}=m_D^T M^{-1} m_D$, we find that the dependence
on the $q^{1}$ charges drops out and only that from $q^{\bar5}$ remains. As a consequence the effective light
neutrino Majorana mass matrix
$m_{\nu}$ can be written  in terms of
$q^{\bar5}$ only: $m_{\nu}=\lambda_{\bar 5} (y_D^T y_M^{-1} y_D) \lambda_{\bar 5}\cdot v_u^2/\Lambda$. 
In addition, for the neutrino mixing matrix $U_{ij}$, which is determined by $m_\nu$ in
the basis where the charged leptons are diagonal, one can prove that 
$U_{ij}\approx \lambda^{|q^{\bar 5}_i-q^{\bar 5}_j|}$, in terms of the differences of the
$\bar5$ charges, when terms that are down by powers of
the small parameter $\lambda$ are neglected. Similarly the CKM matrix elements are approximately determined by only the 10 charges \cite{u1}:
$V^{CKM}_{ij}\approx\lambda^{|q^{10}_i-q^{10}_j|}$.  With these results in mind, we understand that 
the $q^{10}$ charge
assignments in eq. (\ref{cha}) 
are determined by requiring $V_{us}\sim\lambda$, $V_{cb}\sim\lambda^2$ and
$V_{ub}\sim\lambda^3$. However the same $q^{10}$ charges also fix $m_u:m_c:m_t\sim \lambda^6:\lambda^4:1$. The
experimental value of $m_u$ (the relevant mass values are those at the GUT scale: $m=m(M_{GUT})$ \cite{koide}) would rather prefer
$q^{10}_1=4$. Taking into account this indication and the presence of the unknown coefficients $y_{ij}\sim O(1)$ it
is difficult to decide between $q^{10}_1=3$ or $4$ and both are acceptable. 

Turning to the $\bar 5$ charges, the
entries
$q^{\bar 5}_2=q^{\bar 5}_3=0$ have been selected in eq. (\ref{cha}) so that the 22, 23, 32, 33 entries of the effective light neutrino
mass matrix
$m_\nu$ are all O(1) in order to accommodate the nearly maximal value of $s_{23}$. The small non diagonal terms of the
charged lepton mass matrix cannot change this. In fact, for $q^{10}_i$, $q^{\bar 5}_i$ chosen as in eqs. (\ref{cha}) we obtain:   
\beq m_d=
\left(
\matrix{
\lambda^5&\lambda^4&\lambda^2\cr
\lambda^3&\lambda^2&1\cr
\lambda^3&\lambda^2&1}
\right)v_d~~=(m_l)^T~,~~~~m_\nu=
\left(
\matrix{
\lambda^4&\lambda^2&\lambda^2\cr
\lambda^2&1&1\cr
\lambda^2&1&1}\right){v_u^2\over \Lambda}~~~, 
\label{mlep}
\eeq 
where $v_{u,d}$ are the VEVs of the Higgs doublets.
Note that the patterns $m_d:m_s:m_b\sim m_e:m_\mu:m_\tau \sim \lambda^5:\lambda^2:1$ are acceptable (but also
$q^5_1=3$ would be possible). One difficulty is that for $m_\nu$ the subdeterminant 23 is not suppressed in this 
case, so that the splitting between the 2 and 3 light
neutrino masses is in general small. In spite of the fact that $m_D$ is, in first approximation, of the form in eq.
(\ref{md00}) the strong correlations between
$m_D$ and $M$ implied by the simple charge structure of the model destroy the vanishing of the 23 subdeterminant
that would be guaranteed for generic $M$. Models of this sort have been proposed in the literature 
\cite{chargesnu,lopsu1}. The hierarchy between
$m_2$ and
$m_3$ is considered accidental and better be moderate. The preferred solar solution in this case is MSW SA because
if $m_1$ is suppressed (and some suppression is needed if we want $s_{13}$ small) the solar mixing angle is typically
small. However, if with a moderate fine tuning we stretch by hand $m_2$ to become sufficiently close to $m_1$ then
the MSW LA solution could also be reproduced. From eq. (\ref{mlep}), taking $v_u\sim 250~{\rm GeV}$, the mass scale ${\Lambda}$ of
the heavy Majorana neutrinos turns out to be close to the unification scale, 
${\Lambda}\sim 10^{15}~{\rm GeV}$.

A different interesting possibility \cite{mur} is to recover an anarchical picture of neutrinos by taking
$(q^{10}_1,q^{10}_2,q^{10}_3) =  (4,2,0)$ and $(q^{\bar5,1}_1,q^{\bar5,1}_2,q^{\bar5,1}_3) = (0,0,0)$. The 10 charges lead
to an acceptable pattern for the $m_u$ matrix and $V_{CKM}$, as already discussed. For down quarks and charged
leptons we obtain a weakened hierarchy, essentially the square root than that of up quarks: $m_d:m_s:m_b\sim
m_e:m_\mu:m_\tau \sim \lambda^4:\lambda^2:1$. Finally in the neutrino sector the anarchical model is
realized (both $m_D$ and $M$ are structureless).

Note that in all previous cases we could add a constant to $q^{\bar 5}_i$, for example by taking 
$(q^{\bar5}_1,q^{\bar5}_2,q^{\bar5}_3)
=  (4,2,2)$. This would only have the consequence to leave the top quark as the only unsuppressed mass and to
decrease the resulting value of $\tan{\beta}=v_u/v_d$ down to $\lambda^2 m_t/m_b$. A constant shift of the charges $q^1_i$ might also provide a suppression
of the leading $\nu^c$ mass eigenvalue, from $\Lambda_c$ down to the 
appropriate scale $\Lambda$. One
can also consider models where the 5 and $\bar 5$ Higgs charges are different, as in the ``realistic'' SU(5) model of
ref. \cite{afm2}. Also in these models the top mass could be the only one to be non vanishing in the symmetric limit and the
value of $\tan{\beta}$ can be adjusted.

\subsubsection{F(fermions) and F(flavons) of both signs}

Models with naturally large 23 splittings are obtained if we allow negative charges and, at the same time, either
introduce flavons of opposite charges or stipulate that matrix elements with overall negative charge are put to
zero. For example, we can assign to the fermion fields the set of
F charges given by:
\bea
(q^{10}_1,q^{10}_2,q^{10}_3) &= & (3,2,0) \nn\\
(q^{\bar 5}_1,q^{\bar 5}_2,q^{\bar 5}_3) &= & (b,0,0)~~~~~~~~~~~~b\ge 2a>0\nn\\
(q^{1}_1,q^{1}_2,q^{1}_3) &= & (a,-a,0) ~~~.\label{cha1}
\eea   
We consider the Yukawa coupling allowed by U(1)$_{\rm F}$-neutral  Higgs multiplets
in the $5$ and ${\bar 5}$ SU(5) representations and by a pair $\theta$ and
${\bar\theta}$ of SU(5) singlets with F$=1$ and F$=-1$, respectively. 
If $b=2$ or 3, the up, down and charged lepton sectors are
not essentially different than in the previous case. Also in this case the O(1) off-diagonal entry of $m_l$, typical
of lopsided models, gives rise to a large LH  mixing in the 23 block which corresponds to a large
RH mixing in the
$d$ mass matrix. In the neutrino sector, the Dirac and Majorana mass matrices are given by:
\beq m_D=
\left(
\matrix{
\lambda^{a+b}&\lambda^a&\lambda^a\cr
\lambda^{b-a}&{\lambda'}^a&{\lambda'}^a\cr
\lambda^b&1&1}
\right)v_u~~,~~~~~~ M=
\left(
\matrix{
\lambda^{2a}&1&\lambda^a\cr 1&{\lambda'}^{2a}&{\lambda'}^a\cr
\lambda^a&{\lambda'}^a&1}
\right){\Lambda}~~~,
\label{dam}
\eeq 
where $\lambda'$ is given by $\langle\theta\rangle/\Lambda_c$ and ${\Lambda}$ as before denotes the large mass scale associated to the
RH neutrinos: ${\Lambda}\gg v_{u,d}$. After diagonalization of the charged lepton sector and after
integrating out the heavy RH neutrinos we obtain the following neutrino mass matrix in the low-energy
effective theory:
\beq
m_\nu=
\left(
\matrix{
\lambda^{2 b}&\lambda^b&\lambda^b\cr
\lambda^b&1+\lambda^a{\lambda'}^a&1+\lambda^a{\lambda'}^a\cr
\lambda^b&1+\lambda^a{\lambda'}^a&1+\lambda^a{\lambda'}^a}\right)
{v_u^2\over {\Lambda}}~~~.
\label{mnu}
\eeq
The O(1) elements in the 23 block are produced by combining the
large  LH mixing induced by the charged lepton sector and the large LH mixing in $m_D$. A crucial
property of
$m_\nu$ is that, as a result of the see-saw mechanism and of the specific U(1)$_{\rm F}$ charge assignment, the
determinant of the 23 block is automatically of $O(\lambda^a{\lambda'}^a)$ 
(for this the presence of negative
charge values, leading to the presence of both $\lambda$ and
$\lambda'$ is essential \cite{lops2,lopsu5af}). The neutrino mass matrix of eq. 
(\ref{mnu}) is a particular case of the more general pattern 
presented in eq. (\ref{hier1}), for $\delta\approx\lambda^{2b}$,
$\epsilon\approx\lambda^b$ and $\eta\approx\lambda^a{\lambda'}^a$. If we take
$\lambda\approx\lambda'$, it is easy to verify that the eigenvalues of $m_\nu$ satisfy  the relations:
\beq m_1:m_2:m_3  = \lambda^{2(b-a)}:\lambda^{2a}:1~~.
\eeq 
The atmospheric neutrino oscillations require 
$m_3^2\sim 10^{-3}~{\rm eV}^2$. The squared mass difference between the lightest states is  of
$O(\lambda^{4a})~m_3^2$, not far from the MSW solution to the solar neutrino problem if we choose $a=1$. In general $U_{e3}$ is
non-vanishing, of $O(\lambda^b)$. Finally, beyond the large mixing in the 23 sector,
$m_\nu$  provides a mixing angle $\theta_{12} \sim \lambda^{b-2a}$ 
in the 12 sector. For $b> 2a$, we recover a small solar mixing angle. For instance, taking $b=3$ and $a=1$, $\theta_{12}$ becomes close to the range preferred
by the MSW SA solution. When $b=2 a$, as for instance in the case 
$b=2$ and $a=1$, the MSW LA solution can be reproduced. 

\subsubsection{F(fermions) of both signs and F(flavons) $<$0}

A general problem common to all models dealing with flavour is that of recovering the correct  vacuum structure by
minimizing the effective potential of the theory. It may be noticed that the presence of two multiplets $\theta$ and
${\bar \theta}$ with opposite F charges could hardly be reconciled, without adding extra structure to the model,
with a large common VEV for these fields, due to possible analytic terms of the kind $(\theta {\bar \theta})^n$ in
the superpotential. We find therefore instructive to explore the consequences of allowing only the negatively
charged ${\bar \theta}$ field in the theory.

It can be immediately recognized that, while the quark mass matrices 
previously discussed are unchanged, in the
neutrino sector the Dirac and Majorana matrices
are obtained from eq. (\ref{dam}) by setting $\lambda'=0$:
\beq 
m_D=
\left(
\matrix{
\lambda^{a+b}&\lambda^a&\lambda^a\cr
\lambda^{b-a}&0&0\cr
\lambda^b&1&1}
\right)v_u~~,~~~~~~~~ M=
\left(
\matrix{
\lambda^{2a}&1&\lambda^a\cr 1&0&0\cr
\lambda^a&0&1}
\right){\Lambda}~~.
\eeq 
The zeros are due to the analytic property of the superpotential that makes impossible to form the
corresponding F invariant by using ${\bar \theta}$ alone. These zeros should not be taken literally, as they will
be eventually   filled by small terms coming, for instance, from the diagonalization of the charged lepton mass
matrix and from the transformation that put the kinetic terms into canonical form. It is however interesting to work
out, in first approximation, the case  of exactly zero entries in $m_D$ and $M$, when forbidden by F.
The neutrino mass matrix obtained via see-saw from $m_D$ and $M$ has the same pattern as the one displayed in eq.
(\ref{mnu}). A closer inspection reveals that the determinant of the 23 block is identically zero, independently from
$\lambda$. This leads to the following pattern of masses:
\beq 
m_1:m_2:m_3  = \lambda^b:\lambda^b:1~~,~~~~~m_1^2-m_2^2 = {\rm O}(\lambda^{3b})~~.
\eeq 
Moreover the mixing in the 12 sector is almost maximal:
\beq 
\theta_{12}={\pi\over 4}+{\rm O}(\lambda^b)~~.
\eeq 
For $b=3$ and $\lambda\sim 0.2$, both the squared mass difference $(m_1^2-m_2^2)/m_3^2$  and $\sin^2 2\theta_{12}$ are
remarkably close to the values  required by the VO solution to the solar neutrino problem. This
property  remains reasonably stable against the perturbations induced by small terms (of order $\lambda^5$)
replacing the zeros, coming from the diagonalization of the charged lepton sector  and by the transformations that
render the kinetic terms canonical. By choosing $b=2$ we obtain the LOW 
solution. We find quite interesting that also the just-so and the LOW 
solutions, requiring  an
intriguingly small mass difference and a bimaximal mixing, can be described, at least at the level of order of
magnitudes, in the context of a ``minimal'' model of flavour compatible with supersymmetric SU(5). In this case the
role played by supersymmetry  is essential, a non-supersymmetric model with ${\bar \theta}$ alone  not being
distinguishable from the version with both
$\theta$ and ${\bar \theta}$, as far as low-energy flavour properties are concerned.

In conclusion, models based on SU(5) $\times$ U(1)$_{\rm F}$ are clearly toy models that can only aim at a semiquantitative
description of fermion masses. In fact only the order of magnitude of each matrix entry can be specified. However
it is rather impressive that a reasonable description of fermion masses, now also including neutrino masses and
mixings, can be obtained in this simple context, which is suggestive of a deeper relation between gauge and flavour
quantum numbers. There are 12 mass eigenvalues and 6 mixing angles that are specified, modulo coefficients of
order 1, in terms of a bunch of integer numbers (from half a dozen to a dozen), the charges, plus 1 or more scale
parameters. In the neutrino sector we have seen that the scheme is flexible enough to accommodate all the solutions that are still possible. 
Models aiming at a realistic unification of electroweak and strong interactions
should of course address other important questions such as the doublet-triplet
splitting and its stability against quantum corrections, a proton lifetime
compatible with the existing limits, a correct gauge 
coupling constant unification and the consistency with present bounds on 
flavour violation \cite{lfv}. Encompassing all these features in a 
consistent and possibly simple model is a formidable task, that might require
to go beyond the conventional formulation in terms of a four-dimensional
quantum field theory.

\subsection{GUT Models based on SO(10)}

Models based on SO(10) times a flavour symmetry are more difficult to construct because a whole generation is
contained in the 16, so that, for example for U(1)$_{\rm F}$, one would have the same value of the charge for all quarks
and leptons of each generation, which is too rigid. But the mechanism discussed so far, based on asymmetric mass
matrices, can be embedded in an
SO(10) grand-unified theory in a rather economic way
\cite{barr,lops1,soten}. The 33 entries of the fermion mass matrices can be obtained through the coupling
${\bf 16}_3 {\bf 16}_3 {\bf 10}_H$ among the fermions in the third generation, ${\bf 16}_3$, and a Higgs tenplet
${\bf 10}_H$. The two independent VEVs of the tenplet $v_u$ and $v_d$ give mass, respectively, to $t/\nu_\tau$ and
$b/\tau$. The key point to obtain an asymmetric texture is the introduction of an operator of the kind ${\bf 16}_2
{\bf 16}_H {\bf 16}_3 {\bf 16}_H'$ . This operator is thought to arise by integrating out an heavy {\bf 10} that
couples both to ${\bf 16}_2 {\bf 16}_H$ and to ${\bf 16}_3 {\bf 16}_H'$. If the ${\bf 16}_H$ develops a VEV breaking
SO(10) down to SU(5) at a large scale, then, in terms of
SU(5) representations, we get an effective coupling of the kind ${\bf \bar{5}}_2 {\bf 10}_3 {\bf\bar{5}}_H$, with
a coefficient that can be of order one. This coupling contributes to the 23 entry of the down quark mass matrix  and
to the 32 entry of the charged lepton mass matrix, realizing the desired asymmetry.   To distinguish the lepton and
quark sectors one can further introduce  an operator of the form ${\bf 16}_i {\bf 16}_j {\bf 10}_H {\bf 45}_H$,
$(i,j=2,3)$, with the VEV of the 
${\bf 45}_H$ pointing in the $B-L$ direction. Additional operators, still of the type 
${\bf 16}_i {\bf 16}_j {\bf 16}_H {\bf 16}_H'$ can contribute to the matrix elements of the first generation. The
mass matrices look like:
\beq 
m_u=
\left(
\matrix{ 0& 0& 0\cr 0& 0& \epsilon/3\cr 0&-\epsilon/3&1}
\right)v_u~~,~~~~~~~ m_d=
\left(
\matrix{ 0&\delta&\delta'\cr
\delta&0&\sigma+\epsilon/3\cr
\delta'&-\epsilon/3&1}
\right)v_d~~,
\label{mquark1}
\eeq 
\beq m_D=
\left(
\matrix{ 0& 0& 0\cr 0& 0& -\epsilon\cr 0& \epsilon&1}
\right)v_u~~,~~~~~~~ m_l=
\left(
\matrix{ 0&\delta&\delta'\cr
\delta&0&-\epsilon\cr
\delta'&\sigma+\epsilon&1}
\right)v_d~~.
\label{mquark2}
\eeq  
They provide a good fit of the available data in the quarks and the charged lepton sector in terms of 5 
parameters (one of which is complex). In the neutrino sector one obtains a large
$\theta_{23}$ mixing angle,
$\sin^2 2\theta_{12}\sim 6.6\cdot 10^{-3}$ eV$^2$ and $\theta_{13}$ of the same order of
$\theta_{12}$. Mass squared differences are sensitive to the details of the Majorana mass matrix. 

Looking at models with three light neutrinos only, i.e. no sterile neutrinos, from a more general point of view, we
stress that in the above models the atmospheric neutrino mixing is considered large, in the sense of being of order
one in some zeroth order approximation. In other words it corresponds to off-diagonal matrix elements of the same
order of the diagonal ones, although the mixing is not exactly maximal. The idea that all fermion mixings are small
and induced by the observed smallness of the non diagonal $V_{CKM}$  matrix elements is then abandoned. An
alternative is to argue that perhaps what appears to be large is not that large after all. The typical small
parameter that appears in the mass matrices is $\lambda\sim
\sqrt{m_d/m_s}
\sim
\sqrt{m_{\mu}/m_{\tau}}\sim 0.20-0.25$. This small parameter is not so small that it cannot become large due to some
peculiar accidental enhancement: either a coefficient of order 3, or an exponent of the mass ratio which is less
than $1/2$ (due for example to a suitable charge assignment), or the addition in phase of an angle from the
diagonalization of charged leptons and an angle from neutrino mixing. One may like this strategy of producing a
large mixing by stretching small ones if, for example, he/she likes symmetric mass matrices, as from left-right
symmetry at the GUT scale. In left-right symmetric models smallness of left mixings implies that also right-handed
mixings are small, so that all mixings tend to be small. Clearly this set of models \cite{str} tend to favour
moderate hierarchies and a single maximal mixing, so that the MSW SA solution of solar neutrinos is preferred.

\section{Conclusion}

By now there are rather convincing experimental indications for neutrino
oscillations.  The direct implication of these
findings is that neutrino masses are not all vanishing. As a consequence,
the phenomenology of neutrino masses and mixings
is brought to the forefront.  This is a very interesting subject in many
respects. It is a window on the physics of GUTs in
that the extreme smallness of neutrino masses can only be explained in a
natural way if lepton number conservation is
violated.  If so, neutrino masses are inversely proportional to the large
scale where lepton number is violated. Also, the
pattern of neutrino masses and mixings interpreted in a GUT framework can
provide new clues on the long standing problem of
understanding the origin of the hierarchical structure of quark and lepton
mass matrices. Neutrino oscillations only
determine differences of $m_i^2$ values and the actual scale of neutrino
masses remain to be experimentally fixed. In
particular, the scale of neutrino masses is important for cosmology as
neutrinos are candidates for hot dark matter: nearly
degenerate neutrinos with a common mass around 1- 2 eV would significantly
contribute to
$\Omega_m$, the matter density in the universe in units of the critical
density. The detection of
$0\nu\beta\beta$ decay would be extremely important for the determination
of the overall
scale of neutrino masses, the confirmation of their Majorana nature and
the experimental clarification of the ordering of
levels in the associated spectrum. The recent indication of a signal for
$0\nu\beta\beta$ with $m_{ee}$
in a range around 0.4 eV, if confirmed, would point to a small but
possibly non negligible contribution of neutrinos
to
$\Omega_m$ and, among models with 3 neutrinos, would favour those with a
degenerate spectrum. The decay of heavy
right-handed neutrinos with lepton number non-conservation can provide a
viable and attractive model of baryogenesis
through leptogenesis. The measured oscillation frequencies and mixings are
remarkably consistent with this attractive
possibility.

While the existence of oscillations  appears to be on a ground of
increasing solidity, many
important experimental challenges remain. For atmospheric neutrino
oscillations the completion of the K2K experiment, now
stopped by the accident that has seriously damaged the Superkamiokande
detector, is important for a terrestrial
confirmation of the effect and for an independent measurement of the
associated parameters. In the near future the
experimental study of atmospheric neutrinos will continue with long
baseline measurements by MINOS, OPERA, ICARUS.  For
solar neutrinos it is not yet clear which of the solutions, MSW SA, MSW
LA, LOW and VO, is true, although a preference for
the MSW LA solution appears to be indicated by the present data. This issue
will be presumably clarified in the near future by
the continuation of SNO and the forthcoming data from KAMLAND and
Borexino.  Finally a clarification by MINIBOONE of the
issue of the LSND alleged signal is necessary, in order to know if 3 light
neutrinos are sufficient or additional sterile
neutrinos must be introduced, in spite of the apparent lack of independent
evidence in the data for such sterile
neutrinos and of the fact that attempts of constructing plausible and
natural theoretical models have not led so far to
compelling results. Further in the future there are projects for neutrino
factories and/or superbeams aimed at precision
measurements of the oscillation parameters and possibly the detection of
CP violation effects in the neutrino sector.

Pending the solution of the existing experimental ambiguities a large
variety of theoretical models of neutrino
masses and mixings are still conceivable. Among 3-neutrino models we have
described a variety of possibilities based
on degenerate, inverted hierarchy and normal hierarchy type of spectra.
Most models prefer one or the other of the
possible experimental alternatives which are still open. It is interesting
that the MSW LA solar oscillation solution,
which at present appears somewhat favoured by the data, is perhaps the
most constraining for theoretical models. For
example, it is difficult to reproduce this solution in the inverted
hierarchy models. The MSW LA solution can be obtained in
the degenerate case, also including the anarchical scenario, and in the
normal hierarchy case, but also in these cases
rather special conditions must be met. In many cases the MSW LA solution
corresponds to values of $\theta_{13}$ rather large,
not far from the present bound. The values of $m_{ee}$ (which determines
the rate of $0\nu\beta\beta$) that is found in
models leading to the MSW LA solution are typically of order
$\sqrt{\Delta m^2_{sun}}$ in the normal hierarchy case and can be even
larger in the degenerate case, where the expected values are at least of
order $\sqrt{\Delta m^2_{atm}}$ (assuming that indeed for
the MSW LA solution $s_{12}$ is large but not close to maximal).

The fact that some neutrino mixing angles are large and even nearly
maximal, while surprising at the start, was eventually
found to be well compatible with a unified picture of quark and lepton
masses within GUTs. The symmetry group at
$M_{GUT}$ could be either (SUSY) SU(5) or SO(10)  or a larger group. For
example, we have presented a
class of natural models where large right-handed mixings for quarks are
transformed into large left-handed mixings for
leptons by the transposition relation $m_d=m_e^T$ which is approximately
realized in SU(5) models. In particular, we
argued in favour of models with 3 widely split neutrinos. Reconciling
large splittings with large mixing(s) requires some
natural mechanism to implement a vanishing determinant condition. This can
be obtained in the see-saw mechanism, for
example, if one light right-handed neutrino is dominant, or a suitable
texture of the Dirac  matrix is imposed by an
underlying symmetry. We have shown that these mechanisms can be naturally
implemented  by simple assignments of U(1)$_{\rm F}$
horizontal charges that lead to a successful semiquantitative unified
description of all quark and lepton masses in SUSY
SU(5)$\times$ U(1)$_{\rm F}$. Alternative realizations based on the SO(10)
unification group have also been discussed.

In conclusion, the discovery of neutrino oscillations with frequencies
that point to very small $\nu$ masses has opened a
window on the physics beyond the Standard Model at very large energy
scales. The study of the neutrino mass and mixing
matrices, which is still at the beginning, can lead to particularly
exciting insights on the theory at large
energies possibly as large as $M_{GUT}$.
\vspace*{1.0cm}
\section*{Acknowledgements}
We thank J.~Garcia-Bellido, A. Masiero, I. Masina, A. Riotto, A. Strumia and F. Vissani 
for discussions. F.F. thanks the CERN Theoretical Division, where part of this work was 
done, for hospitality and financial support. F.F. is partially
supported by the European Programs HPRN-CT-2000-00148 and HPRN-CT-2000-00149.
\newpage
%

%
\end{document}